\documentclass[letterpaper,twocolumn,10pt]{article}
\usepackage{zhanggroup}

\usepackage{tikz}
\usepackage{amsmath}
\usepackage{filecontents}
\usepackage{url}
\usepackage{footnote}
\usepackage{mathtools}
\usepackage{amsthm}
\usepackage{amssymb}
\usepackage{dsfont}
\usepackage{pifont}
\usepackage{bbm}
\usepackage{caption}
\usepackage{subcaption}
\usepackage{float}
\usepackage{tabularx}
\usepackage{makecell}
\usepackage{booktabs}
\usepackage{mathtools}
\usepackage{multirow}
\usepackage{xcolor}
\usepackage{colortbl}
\usepackage{enumitem}
\usepackage[flushleft]{threeparttable}
\usepackage[ruled, vlined, linesnumbered]{algorithm2e}
\usepackage[pagebackref=true,breaklinks=true,colorlinks,bookmarks=false]{hyperref}
\usepackage[sort&compress,numbers]{natbib}
\usepackage[absolute]{textpos}
\hypersetup{
  colorlinks,
  linkcolor={blue!60!green},
  citecolor={green!60!blue},
  urlcolor={orange!60!red}
}
\DeclareMathOperator*{\argmax}{arg\,max}
\DeclareMathOperator*{\argmin}{arg\,min}
\DeclarePairedDelimiter\ceil{\lceil}{\rceil}

\newcommand{\mypara}[1]{\noindent{\bf {#1.}}}
\newcommand{\pair}[1]{\langle {#1} \rangle}
\newcommand{\teachermodel}{\mathcal{T}}
\newcommand{\studentmodel}{\mathcal{S}}

\newcommand{\featureextractor}{\mathcal{F}_{\mathtt{T}}}
\newcommand{\featureextractorNumbered}[1]{\mathcal{F}_{\mathcal{T}_{#1}}}
\newcommand{\classifier}{\mathcal{C}_{\mathtt{S}}}
\newcommand{\dataset}{\mathcal{D}}
\newcommand{\teacherdataset}{\dataset_{\mathtt{T}}}
\newcommand{\studentdataset}{\dataset_{\mathtt{S}}}
\newcommand{\modelinput}{\mathbf{x}}
\newcommand{\newvar}{\mathbf{w}}
\newcommand{\zerovar}{\mathbf{0}}
\newcommand{\modeloutput}{y}
\newcommand{\inputspace}{\mathcal{X}}

\newcommand{\prob}[1]{\text{Pr}\left({#1}\right)}
\newcommand{\blackbox}{f}
\newcommand{\matchset}{S_{\textup{match}}}
\newcommand{\supportset}{S_{\textup{support}}}

\newcommand{\threshold}{\tau}
\newcommand{\proportion}{P_{\textup{match}}}
\newcommand{\fingerprintingvector}{v_{\textup{fgpt}}}
\newcommand{\eccentricity}{E}
\newcommand{\entropy}{H}
\newcommand{\indicator}{\mathds{1}}
\newcommand{\softmax}[1]{\texttt{SF}(#1)}
\newcommand{\fullyconnected}[1]{\texttt{FC}(#1)}

\newcommand{\batchnorm}{\texttt{BN}}
\newcommand{\sigmoid}[1]{\texttt{SG}(#1)}
\newcommand{\cmark}{\ding{51}}
\newcommand{\xmark}{\ding{55}}
\newtheorem{definition}{Definition}[section]
\newtheorem{theorem}{Theorem}
\newtheorem*{remark}{Remark}

\begin{document}
%==================================================

%==================================================
\begin{textblock}{12}(2,1)
\centering
To Appear in the 31st USENIX Security Symposium, August 10–12, 2022.
\end{textblock}

\date{}
%==================================================

%==================================================
\title{\Large \bf Teacher Model Fingerprinting Attacks Against Transfer Learning}
%==================================================

%==================================================
\author{
{\rm Yufei Chen\textsuperscript{1,2}}\ \ \ \ \ 
{\rm Chao Shen\textsuperscript{1}}\ \ \ \ \
{\rm Cong Wang\textsuperscript{2}}\ \ \ \ \ \
{\rm Yang Zhang\textsuperscript{3}}
\\
\\
\textsuperscript{1}\textit{Xi'an Jiaotong University}\ \ \
\textsuperscript{2}\textit{City University of Hong Kong}\\
\textsuperscript{3}\textit{CISPA Helmholtz Center for Information Security}
}
%==================================================

%==================================================
\maketitle
%==================================================

%================================================== 
\begin{abstract}
%================================================== 

Transfer learning has become a common solution to address training data scarcity in practice.
It trains a specified student model by reusing or fine-tuning early layers of a well-trained teacher model that is usually publicly available.
However, besides utility improvement, the transferred public knowledge also brings potential threats to model confidentiality, and even further raises other security and privacy issues. 
    
In this paper, we present the first comprehensive investigation of the teacher model exposure threat in the transfer learning context, aiming to gain a deeper insight into the tension between public knowledge and model confidentiality.
To this end, we propose a \textit{teacher model fingerprinting attack} to infer the origin of a student model, i.e., the teacher model it transfers from.
Specifically, we propose a novel optimization-based method to carefully generate queries to probe the student model to realize our attack.
Unlike existing model reverse engineering approaches, our proposed fingerprinting method neither relies on fine-grained model outputs, e.g., posteriors, nor auxiliary information of the model architecture or training dataset.
We systematically evaluate the effectiveness of our proposed attack.
The empirical results demonstrate that our attack can accurately identify the model origin with few probing queries.
Moreover, we show that the proposed attack can serve as a stepping stone to facilitating other attacks against machine learning models, such as model stealing.\footnote{Our code is available at \url{https://github.com/yfchen1994/Teacher-Fingerprinting}.}

%================================================== 
\end{abstract}
%================================================== 

%==================================================
\section{Introduction}
%==================================================

The past decade has witnessed an unprecedented development of machine learning (ML).
Yet, the progress of ML heavily relies on sophisticated models, sufficient computing resources, and a massive volume of training data, which remain major constraints to building high-performance ML models.

Transfer learning opens a pathway for overcoming obstacles raised by the lack of data or computing resources; it is essentially an ML paradigm to transfer the knowledge from a well-learned domain into a specified domain where training data are scarce.
Concretely, transfer learning establishes a new model (a.k.a. \textit{student model}) through borrowing early layers from a pre-trained model (a.k.a. \textit{teacher model}), which requires far less efforts than training from scratch.
It has been proved to be a promising ML practice and widely applied in a wide range of academic and industrial areas, such as computer vision~\cite{LCWJ15}, natural language processing~\cite{ZYMK16}, etc.

Despite the huge success, such cross-domain knowledge learning paradigm also raises security and privacy concerns:
\begin{itemize}[noitemsep]
\item There exist many advanced ML models, each of which can potentially serve as a teacher model for transfer learning.
Choosing an appropriate teacher model to train a student model requires a large number of engineering efforts.
Thus, for a student model, the choice of its teacher model certainly belongs to the model owner's intellectual property (IP), and should be kept confidential.
\item On the other side of the coin, from the responsible ML perspective, a teacher model owner does not want their model to be transferred to perform unethical or illegal tasks, such as facial recognition or weapon classification.
Thus, a teacher model owner needs a way to track the parties that use their model to build student models.
\item Malicious parties can intentionally publish vulnerable ML models online.
When such models are used as teacher models for transfer learning, the corresponding student models may inherit some vulnerabilities.
For instance, Zhang et al. demonstrate that fine-tuned transfer learning models are more prone to transferable adversarial examples~\cite{ZSLBY20}.
Yao et al. show that an attacker can infect transfer learning models by distributing pre-trained models with backdoors~\cite{YLZZ19}.
\item Also, after learning the teacher model that a student is transferred from, an attacker can perform more effective malicious attacks, such as adversarial example attacks~\cite{CW17} and model stealing/extraction attacks~\cite{TZJRR16}. 
\end{itemize}

All these concerns require us to gain a deeper insight into the teacher model exposure or transfer learning.

In this paper, we present the first comprehensive investigation of the teacher model exposure threat in the transfer learning context. 
In particular, we propose a novel teacher model fingerprinting attack that can effectively identify the teacher model of a transferred student.
The key idea is to generate a set of \textit{fingerprinting pairs} for each teacher model candidate, which will activate similar latent feature representations.
Hopefully, most of them will also trigger similar latent features and bring a pair of similar responses to the student model, if the student model is transferred from the teacher model candidate.
We formalize the query generation process as an optimization problem, which can be solved via gradient descent.
Note that, the fingerprinting pairs for each teacher model can be applied to all student models transferred from the teacher model, i.e., they are reusable.
This point demonstrates the efficiency and practicality of our method.

We conduct extensive experiments to investigate the effectiveness of our proposed method, and thoroughly evaluate how various factors affect attack performance.
The evaluation shows that our attack can achieve high teacher model inference accuracy even with a limited number of queries in top-1 label exposure. 
Moreover, our attack still works well when synthesizing queries from samples unrelated to the target domain, or even from random noises.
In addition, we show that our fingerprinting attack can serve as a stepping stone to performing more effective model stealing attacks.

Our study should not be confused with recent research efforts on model stealing or model reverse engineering:

Their primary goal is to directly steal the exact victim model, build a surrogate model with similar functionality, or infer other internal model information.
By contrast, our proposed attack essentially focuses on exploiting the model sharing practice in transfer learning, to quickly identify the presence of pre-trained components.
Once identifying the teacher model components, the attacker has stepped further to realize the aforementioned malicious or unethical goals. 

Besides, from the technical perspective, our attack remedies the key weakness—significance attack expense caused by massive attack queries or shadow models—of prior arts.
Our attack just needs to identify the transferred feature map with few queries, rather than recovering model parameters or architectures interactively.

Furthermore, previous studies are subject to extra attack assumptions, such as transparent model architectures~\cite{CJM20} and fine-grained model outputs~\cite{TZJRR16}.
E.g., the teacher model fingerprinting method proposed by Wang et al.~\cite{WYVZZ18} requires access to posteriors of the target model prediction.
However, in the real world, raw model outputs are usually perturbed or hidden for security~\cite{JSBZG19} or other product deployment concerns.
Instead, our work is set up on a more generic and realistic scenario.
We assume the attacker exclusively receives the victim black-box model's final decision, i.e., the top-1 classification result, which gives the minimum amount of information.
This assumption implies that our proposed attack is likely to be mounted on any transfer learning classifiers.
Additionally, we assume the attacker has zero knowledge of the victim's training dataset, but they are allowed to utilize public data.
These assumptions ensure our proposed method is feasible and applicable in practice.

Our main contributions can be summarized as follows:
\setlist{nolistsep}
\begin{itemize}
\item We take the first step to comprehensively investigate the teacher model exposure in the transfer learning context. In particular, we demonstrate a teacher model fingerprinting attack against transfer learning.
\item We propose a novel optimization-based technique to implement our attack.
\item Extensive evaluations demonstrate the efficacy of our method, and we further show that our teacher model fingerprinting attack can be used to facilitate model stealing attacks against ML models.
\end{itemize}

%================================================== 
\section{Preliminaries}
%================================================== 

%==================================================
\subsection{Problem Statement}
%==================================================

\mypara{Transfer learning}
Transfer learning is commonly used to solve the data-hungry problem of ML, especially for deep learning models~\cite{KBZPYGH20,SMOGSF18,PNIGCLZ18}.
The core is to leverage the feature maps learned by the teacher model, so as to avoid significant cost of training a large model from scratch.

In the source domain, the teacher model $\teachermodel$ has been trained on a large scale dataset $\teacherdataset=\{(\modelinput_i,\modeloutput_i)\}_{i=1}^{N_\mathtt{T}}$,\footnote{In this paper, the notation $\{\cdot\}$ refers to a \textit{multiset}, which may contain repeated values.}
and then it is released for other downstream ML developers.
In a typical transfer learning workflow, the student model $\studentmodel$ is firstly constructed with early $k$ layers copied from $\teachermodel$ for feature extraction, as well as newly added classification layers $\classifier(\cdot)$ (e.g., fully connected layers, SVM, etc.) on the top.
Then, $\studentmodel$ gets trained on a usually confidential dataset $\studentdataset=\{\modelinput_j,\modeloutput_j)\}_{j=1}^{N_\mathtt{S}}$, where $N_{\mathtt{S}} \ll N_{\mathtt{T}}$.
In practice, there are two training strategies:
\begin{itemize}
\item \textbf{Feature extractor}, where all pre-trained layers are frozen to compose a feature extractor $\featureextractor$. 
\item \textbf{Fine-tuning}, where part of the pre-trained parameters of $\featureextractor$ will be updated for better fitting on $\studentdataset$.
\end{itemize}
After being carefully trained, $\studentmodel$ is deployed into a Machine-Learning-as-a-service (MLaaS) platform to provide services.
A customer can send a query $\modelinput$ to the MLaaS service and get the corresponding result $f(\studentmodel(\modelinput))$, via an authorized API $\blackbox(\cdot)$.

\mypara{Teacher model fingerprinting}
In this paper, we develop a novel approach to infer the origin of a student model with only black-box access.
Our motivation is that the student model is likely to inherit fingerprintable model behaviors, i.e., the way to extract features in our case (a.k.a. \textit{feature map}), from the teacher model. 
Some attackers can possibly elicit such fingerprintable behaviors by sending carefully crafted queries to the student model, which is sometimes concealed into an MLaaS platform.
Consequently, they are able to identify the teacher model based on the victim's responses.
We refer to this process as \textit{teacher model fingerprinting attacks}.

%==================================================
\subsection{Threat Model}
%==================================================

\mypara{Attack motivations}
We first list potential motivations to conduct the teacher model fingerprinting attack.

\noindent\textit{Breaking model confidentiality.}
Internal model setup, such as architecture or parameters, is supposed to be kept confidential for the sake of IP protection or security consideration.
However, the model behind an MLaaS platform is no longer a secret once the teacher model has been identified.
The attacker can effortlessly recover early layers with the publicly transparent teacher model information.

\noindent\textit{Stepping stone to advanced attacks.}
As mentioned before, our teacher model fingerprinting attack provides a cheap way to recover most parameters inherited from the teacher model, which helps to open the ``black box.''
Once opening the black box, the attacker has stepped further to discover and exploit vulnerabilities of the student model. 
For instance, they can easily conduct various white-box adversarial attacks against the victim model, such as adversarial example attacks~\cite{CW172} and membership inference attacks~\cite{LF20}.
Also, recent studies have shown that transfer learning may transfer vulnerabilities from the teacher~\cite{ZSLBY20}, which the attacker can directly exploit.  

\noindent\textit{Forensics.}
Despite the aforementioned malicious goals, we can also treat our teacher model fingerprinting attack as a new forensic tool for ML applications. 
For instance, a teacher model owner does not want their model to be transferred to perform unethical or illegal tasks, such as weapon classification.
A potential IP protection solution is to embed watermarks to model~\cite{ABCPK18,ZGJWSHM18,LHZG19}.
However, we are witnessing an arms race right now, where watermarking schemes would be broken by new attacks~\cite{MPT17}.
Our fingerprinting approach can complement watermarking for IP protection.
Also, in some cases, certain parties can intentionally publish a vulnerable teacher model, which will result in the corresponding student models inheriting vulnerabilities~\cite{YLZZ19}.
For both of these cases, our attack can be used by a third-party forensic service as a defense to track the origin of a student model.

\mypara{Threat model in detail}
Then, we describe the detailed threat model in real-world settings.
For our teacher model fingerprinting attack, we consider an attacker with black-box access to the victim student model $\studentmodel$.
The attacker is able to send an arbitrary input $\modelinput$ to $\studentmodel$ via an authorized API $\blackbox(\cdot)$ and receive the corresponding response $\blackbox(\studentmodel(\modelinput))$.
Three cases may arise.

\setlist{nolistsep}
\begin{itemize}[noitemsep]
\item \mypara{Case 1}
$\blackbox(\studentmodel(\modelinput))=\studentmodel(\modelinput)$. In a low-level security setting, the response exposes the exact model output, which contains class labels and raw confidence values.  
\item \mypara{Case 2}
$\blackbox(\studentmodel(\modelinput))=\studentmodel(\modelinput)+\epsilon$. The MLaaS provider returns a perturbed version of the model output to avoid privacy breaches like membership inference attacks~\cite{JSBZG19}.
It is notable that the perturbation should not change the top-1 predicted label~\cite{JSBZG19}, i.e., $\argmax_i \left(\studentmodel(\modelinput)+\epsilon\right)_i == \argmax_i \studentmodel(\modelinput)_i$ ( $S(\modelinput)_i$: the $i$-th component of $S(\modelinput)$).
\item \mypara{Case 3}
$\blackbox(\studentmodel(\modelinput))=\argmax_i \studentmodel(\modelinput)_i$. 
Only the top-1 label is returned, giving the minimal piece of information~\cite{LZ21}.
\end{itemize}

This work demonstrates the teacher model fingerprinting attack in the most restrictive case---top-1 label exposure, i.e., Case 3 where $\blackbox(\modelinput)=\argmax_i \studentmodel(\modelinput)_i$.\footnote{In the remaining, we abuse the notation $\blackbox(\modelinput)$ to represent $\blackbox(\studentmodel(\modelinput))$.}
In this case, our proposed attack is compatible with Case 2: on the one hand, the output perturbation in Case 2 would not change the top-1 predicted label;
one the other hand, even the perturbed model output at least provides information no less than Case 3. 

In our basic setup, we assume the attacker has obtained most potential teacher model candidates $\{ \teachermodel_i \}$ from public resources (e.g., ML frameworks like PyTorch, or websites like ModelZoo~\cite{ModelZoo}). 
We also consider possible cases where the victim model does not come from one of the candidate teacher models, or it is not trained through transfer learning.
Neither the teacher dataset $\teacherdataset$ in the source domain nor the student dataset $\studentdataset$ in the target domain is available.
However, the attacker can collect public datasets like ImageNet to help perform the teacher model fingerprinting attack.
The attacker has two primary goals: 
one is to infer the teacher model accurately,
while the other is to use as few probing queries as possible, to limit attack costs and keep the attack stealthy.

%================================================== 
\section{Teacher Model Fingerprinting}
%==================================================

%==================================================
\subsection{Overview}
%==================================================

In the beginning, we formalize the teacher model fingerprinting attack in the transfer learning context:
\begin{definition}[Teacher Model Fingerprinting Attack]
Suppose there is an attacker given a set of $N$ realistic inputs $\{\modelinput_i\}$ (a.k.a. \textit{probing input}), a set of teacher model candidates $\{\teachermodel_j\}$, and an authorized API $\blackbox(\cdot)$ to the target black-box student model $\studentmodel$ in only top-1 label exposure, the teacher model fingerprinting attack is to infer which $\teachermodel_j$ is adopted by $\studentmodel$.
\end{definition}

\begin{figure}[!t]
\centering
\includegraphics[width=\columnwidth]{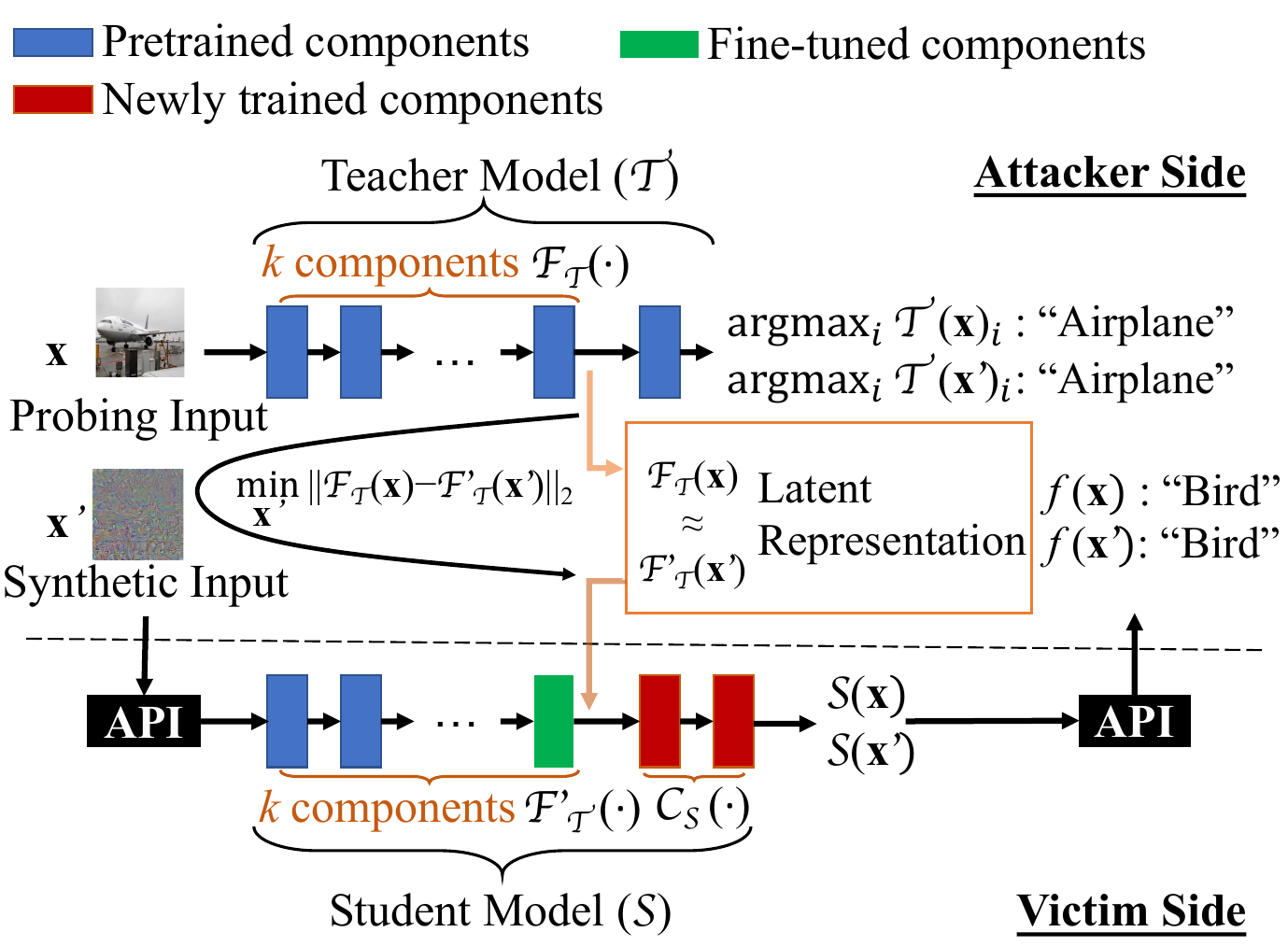}
\caption{
Illustration of our fingerprinting attack.
Given a probing input $\modelinput$ and a teacher model $\teachermodel$, the attacker aims to find a synthetic input $\modelinput'$, which will activate a similar latent representation at the $k$-th component with $\modelinput$: $\featureextractor(\modelinput') \approx \featureextractor(\modelinput)$.
If the first $k$ components of the student model $\studentmodel$ comes from the teacher model $\teachermodel$, some of which may get slightly fine-tuned (still, $\featureextractor'(\cdot)\approx\featureextractor(\cdot)$), the student model is likely to produce a pair of similar responses $\studentmodel(\modelinput') \approx \studentmodel(\modelinput)$.
}
\label{fig:fingerprinting_illustration}
\end{figure}

We start from the feature extractor transfer learning setting.
Suppose there is a student model $\studentmodel$ composed of a feature extractor $\featureextractor$, with early $k$ components copied from the teacher model $\teachermodel$, and newly trained classification layers $\classifier$.
Our intuition is that given a \textit{probing input} $\modelinput$, we can artificially craft a \textit{synthetic input} $\modelinput'$ so that $\featureextractor(\modelinput')\approx\featureextractor(\modelinput)$.
Hopefully, we will activate similar outputs on the student model as $\studentmodel(\modelinput')=\classifier(\featureextractor(\modelinput'))\approx\classifier(\featureextractor(\modelinput))=\studentmodel(\modelinput)$, and therefore have a greater chance to receive similar responses from the API.
In this case, we refer to the pair $\pair{\modelinput, \modelinput'}$ as a \textit{fingerprinting pair}.
Although the above intuition stands on the feature extractor scenario, in~\autoref{section:fine_tuned_extractor}, we will show that our attack still works in the fine-tuning setting. 

\autoref{fig:fingerprinting_illustration} provides a schematic view of our teacher model fingerprinting attack.
Overall, it follows two steps:

\mypara{Step I: generating a set of fingerprinting pairs $\{\pair{\modelinput_i, \modelinput_i'}\}$} 
To perform the teacher model fingerprinting attack, the attacker firstly collects a set of candidate teacher models and realistic probing images.
For each candidate teacher model $\teachermodel_j$, the attacker needs to determine which components may constitute the feature extractor $\featureextractorNumbered{j}$ for the student model.
Conventionally, the student model will adopt the whole convolution part from the teacher model.
Then, for each probing input $\modelinput_i$, the attacker aims to craft a synthetic input $\modelinput_i'$, which will trigger similar latent representations on $\featureextractorNumbered{j}$.
We will study how the choice of components of $\featureextractorNumbered{j}$, the number of $\{\modelinput_i\}$, as well as the source of $\{\modelinput_i\}$ affect the proposed attack in~\autoref{section:probing_features}, \autoref{section:query_budget}, and \autoref{section:inference_resource}, respectively.

This attack query generation procedure ensures our attack is a cheap one, since:
i) the query generation procedure is locally conducted without any expense by the target MLaaS service, and
ii) for fingerprinting pairs bonded to $\featureextractorNumbered{j}$, they are reusable to probe whether other student models come from $\featureextractorNumbered{j}$, without further computation costs on query generation.

\mypara{Step II: inferring the teacher model according to black-box responses $\{\pair{\modeloutput_i, \modeloutput_i'}\}$}
After obtaining fingerprinting pairs, the next step is to send them to the target black box, and infer which teacher model is used according to the responses.
Ideally, if the target model relies on the feature extractor $\mathcal{F}_{\mathcal{T}_j}$, most generated fingerprinting pairs would produce a pair of responses matched on the same label.
The inference stage follows the ``one-of-the-best'' strategy: choosing the candidate owning the most matched responses as the inference result.
In~\autoref{section:statistical_analysis}, we will introduce a strategy for robust inferences.

%==================================================
\subsection{Step I: Attack Query Generation}
%==================================================

For our teacher model fingerprinting attack, we use the $L_2$-norm metric to measure the similarity of latent representations.
Given a probing input $\modelinput$ and a candidate feature extractor $\featureextractor$, we formalize the attack query generation task as
\begin{equation}
\label{equ:query_generation}
\begin{aligned}
&\modelinput'=\argmin_{\tilde{\modelinput}}\|\featureextractor(\tilde{\modelinput})-\featureextractor(\modelinput)\|_2,
\text{~s.t.~} \tilde{\modelinput}\in[0,255].
\end{aligned}
\end{equation}

For the constrained optimization problem~\autoref{equ:query_generation}, we adopt the optimization strategy proposed by Carlini and Wagner~\cite{CW17}:
by introducing a new variable $\newvar$, and setting
\begin{equation}
    \tanh(\newvar)=\frac{2\tilde{\modelinput}}{255}-1,
\end{equation}
we have $-1 \leq \tanh(\newvar) \leq 1$, and
\begin{equation}
\begin{aligned}
    -1 \leq \frac{2\tilde{\modelinput}}{255}-1 \leq 1 
    \Rightarrow& 0 \leq \tilde{\modelinput} \leq 255,
\end{aligned}
\end{equation}
which satisfies the constraints.
Therefore, the original problem~\autoref{equ:query_generation} can be converted to
\begin{equation}
\begin{aligned}
    \newvar' = \argmin_\newvar \left\| \featureextractor\left(\frac{255}{2}\left(\tanh({\newvar})+1\right)\right) - \featureextractor(\modelinput) \right\|_2,
\end{aligned}
\end{equation}
which can be solved by gradient descent algorithms. 
We use the Adam optimizer~\cite{KB15}, with learning rate set as $10^{-3}$ for 30,000 iterations.
In our implementation, we initialize $\newvar$ as $\zerovar$.

However, due to type casting (floating points to integers) and optimization loss, $\featureextractor(\modelinput)$ cannot be perfectly equal to $\featureextractor(\modelinput)$.
So after obtaining the synthetic input $\modelinput'$, we will examine whether $\argmax_i\teachermodel(\modelinput)_i==\argmax_j\teachermodel(\modelinput')_j$ on the candidate.
If not, we will discard $\modelinput'$ and generate a new one.

%==================================================
\subsection{Step II: Teacher Model Inference}
%==================================================

The next step is to infer whether the student model comes from one of the candidate teacher feature extractors.
For each candidate feature extractor, the attacker can generate a set of fingerprinting pairs and send them to the black-box student model.
Intuitively, for a specific feature extractor $\featureextractor$, the more matched pairs of black-box responses are obtained, the more likely the target model is transferred from $\featureextractor$.
Here we define the \textit{matching set} useful for further discussion.

\begin{definition}[Matching Set]
After sending $N$ fingerprinting pairs, all the pairs triggering two same API responses ($\modeloutput_i==\modeloutput_i')$ compose the matching set $\matchset$.
\end{definition}

We use three heuristics to measure how much ``belief'' the attacker has to infer the teacher feature extractor.

\mypara{Matching proportion}
A high matching proportion of fingerprinting pairs indicates a high possibility that the target student model uses the same feature extractor as the attacker. 
We compute the matching proportion as a primary metric to depict how ``perfectly matched'' the candidate is with the target feature extractor.
\[
    \proportion=\frac{|\matchset|}{N}.
\]
The attacker will obtain a fingerprinting vector composed of matching proportions $\fingerprintingvector=[\proportion^1, \proportion^2, \cdots]$ for all candidate feature extractors $[\featureextractorNumbered{1}, \featureextractorNumbered{2}, \cdots]$.
Then the attacker will choose the candidate feature extractor that achieves the highest $\proportion$ as their inference result.
Moreover, there are cases where the actual teacher model does not belong to the candidate feature extractor set, or the target model is not trained by transfer learning. 
To handle these cases, our attack introduces a pre-defined threshold $\threshold$.
If $\max(\fingerprintingvector) < \threshold$, the inference result will be set as \texttt{NULL}, which means that the target model is not transferred from the candidate feature extractors.

A high matching proportion is insufficient to prove that the candidate feature extractor is adopted by the student model. 
It is also possible to obtain matched pairs from unmatched features.
That is, the latent features extracted from probing inputs and synthetic inputs are significantly different.
We will dive into such ``false matching'' phenomenon in~\autoref{section:statistical_analysis}.

\mypara{Eccentricity}
Eccentricity is adopted by \cite{NS09} to measure how an item ``stands out'' from the rest in a set.
It is defined as
\[
    \eccentricity(v)=\frac{\text{max}(v)-\text{max}_2(v)}{\sigma(v)},
\]
where $\text{max}(\cdot)$ and $\text{max}_2(\cdot)$ refers to the first and second highest value, respectively, and $\sigma(\cdot)$ refers to the standard deviation.
The higher the eccentricity of $\fingerprintingvector$, the more distinguishing is $\text{max}(v)$, and the more confident the attacker is to make the inference.

\mypara{Empirical entropy}
Here, we introduce the empirical entropy of a set as a heuristic to measure how much ``information'' the set presents. 
Formally, for a set $\{x_1, x_2, \cdots, x_n\}$ and a sample space $\mathcal{X}$, the empirical entropy is defined as:
\[
    \entropy(\mathcal{X})=-\sum_{x \in \mathcal{X}}\hat{p}(x)\log\hat{p}(x),
\]
where
\[
    \hat{p}(x) = \frac{1}{n}\sum_{i=1}^{n}\indicator(x_i==x),
\]
and $\indicator(\cdot)$ is the indicator function.
We calculate the empirical entropy of the \textbf{matching set} with the highest $\proportion$, to estimate how much information is given to make the inference.

%================================================== 
\section{Evaluation}
\label{section:evaluation}
%==================================================

%==================================================
\subsection{Dataset}
%==================================================

We use the following datasets to train student models:

\mypara{Dogs-vs-Cats}
Our Dogs-vs-Cats dataset contains 12,500 dog and 12,500 cat images from the Internet~\cite{Dogs_vs_Cats}.
We select the first 10,000 dog images and the first 10,000 cat images to compose the training set. 
The remaining 2,500 dog images and 2,500 cat images compose the testing set.

\mypara{MNIST}
The MNIST dataset~\cite{MNIST} is a widely used dataset to build up toy image recognition ML models.
It consists of 60,000 training and 10,000 testing handwritten digit samples, containing ten classes from digit 0 to digit 9.

\mypara{CIFAR10 and CIFAR100} 
The CIFAR10 and CIFAR100 datasets~\cite{CIFAR} are another two widely adopted ML datasets.
The CIFAR10 dataset contains ten classes, with 5,000 training samples and 1,000 test samples in each.
The CIFAR100 dataset is similar to the CIFAR10, except that the former has 100 classes with 600 images in each, including 500 training and 100 testing images.

\mypara{STL10}
The STL10 dataset~\cite{STL10} is originally designed to develop supervised as well as unsupervised learning models. 
Our experiment uses the labeled set covering ten classes, with 500 training images and 800 test images per class.

\mypara{CelebA}
We use the first 50,000 facial images as training samples and the following 10,000 facial images as the test samples from the CelebA dataset~\cite{LLWT15}, which are annotated with 40 binary attributes.
For CelebA, we build up multi-label transfer learning models to tag each image with 40 attributes.

We also use different datasets to generate probing inputs:

\mypara{VOC-Segmentation}
Our experiments assume the attacker has obtained the training dataset for the segmentation task of the VOC2012 challenge~\cite{VOC2012}, consisting of 1,464 samples.
Besides, we further assume the image annotation information (i.e., segmentation and class information) is unavailable.

\mypara{Random Noise}
This dataset simulates the case that the attacker cannot acquire any realistic image to build up the attack dataset. 
In this scenario, they have to synthesize images with randomly generated pixels to compose the attack dataset.
We assume the attacker samples each pixel by the normal distribution and normalizes it into an integer within $[0,255]$.

For all images used in our experiments, they are firstly preprocessed into the 8-bit $224\times224$ RGB format.

%==================================================
\subsection{Experiment Setup}
%==================================================

\mypara{Teacher models}
We have downloaded pre-trained models AlexNet, DenseNet121, MobileNetV2, ResNet18, VGG16, VGG19, and GoogLeNet from the official repository of PyTorch.
Furthermore, to examine whether our proposed attack can discriminate teacher feature extractors with the same architecture trained by different organizations, we have also downloaded an AlexNet model from the PyTorchCV package repository~\cite{pytorchcv}.
The detailed information of the pre-trained models is listed in~\autoref{table:teacher_model}.
We adopt the whole convolution part of each pre-trained model as the feature extractor in our experiments.
Particularly, for VGG16 and VGG19, we also adopt the fully connected layers, except the last output layer. 

\mypara{Student models}
For the Dog-vs-Cats, MNIST, STL10, CIFAR10, and CIFAR100 dataset, we develop multi-class classification students models. 
As for the CelebA dataset, we develop multi-label models which annotate the input with 40 binary attributes simultaneously.
We build up three student models for each pre-trained feature extractor individually, by appending different fully connected layers.
The detailed transfer learning setup is described in~\autoref{sec:transfer_learning_setup}.
Our basic transfer learning setup is the \textit{feature extractor} approach.
That is, the parameters of the pre-trained feature extractor are fixed.
Furthermore, in~\autoref{section:fine_tuned_extractor}, we will evaluate the attack performance when some components of the feature extractor get fine-tuned during the transfer learning process.

\mypara{Other targets not trained with transfer learning}
In our experiments, we also consider target models trained from scratch, which have the same model architectures with AlexNet and ResNet18 student models.

\mypara{Basic setup}
Here we introduce the basic attack setup, followed by most of our experiments unless otherwise specified.
We assume the attacker has owned seven pre-trained feature extractors from public resources (models listed by~\autoref{table:teacher_model} except for GoogLeNet), and they have crafted 100 fingerprinting pairs for each teacher feature extractor.
For the ease of comparison, we also assume that for each candidate teacher model, the attacker knows how many layers of the teacher model will be used (e.g., $k$ in~\autoref{fig:fingerprinting_illustration} is known).
We will investigate the case when $k$ is unknown in~\autoref{section:probing_features}.
We randomly select the attacker's probing images from the VOC-segmentation dataset, which does not overlap with the training dataset of the teacher models or student models.
In~\autoref{section:inference_resource}, we will investigate the attack effectiveness when no realistic probing image is available.
Only the top-1 classification label will be reported by the black-box target, except that the multi-label classifier trained on the CelebA dataset will return 40 facial attributes.
We assume the input format and input pre-processing module are transparent to the attacker.
For most real-world applications, the input format is described by the API reference book.
Otherwise, it is also feasible to infer the input format efficiently with some reverse engineering techniques~\cite{XCSCL19}.

\mypara{Choice of $\threshold$}
We assume our target is a $c$-class classifier.
When the classifier simply outputs random results, the probability that the attacker receives a pair of matched responses is $\frac{1}{c}$.
To avoid such random matching, we should at least ensure $\threshold > \frac{1}{c}$.
A generic way is to preset a positive number $\beta>1$ and let $\threshold=\frac{\beta}{c}$.
Also, we must ensure $\beta\leq2$ so as to when $c=2$, $\threshold$ will not exceed 1.
When there are no auxiliary dataset and model to help determine $\tau$, we directly set $\tau=\frac{1.5}{c}$. 

\begin{figure}
\centering
\includegraphics[width=.185\textwidth]{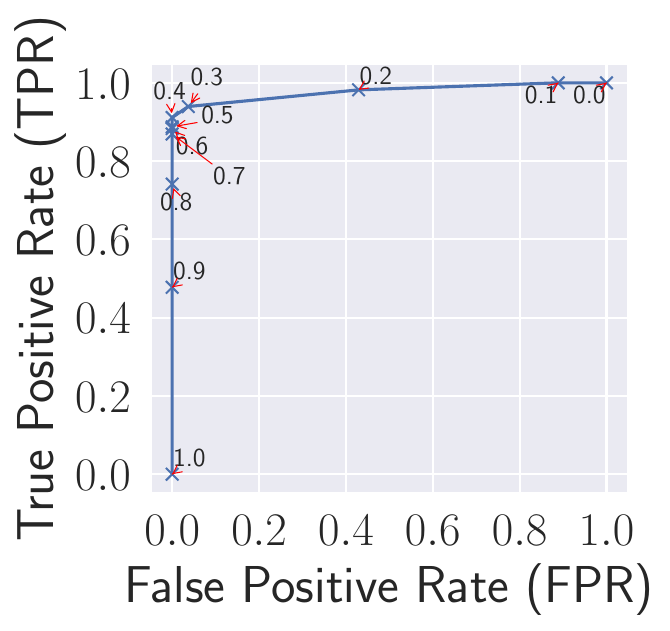}
\caption{
ROC curve with different $\threshold$. Each point refers to a $\threshold$.
}
\label{figure:threshold}
\end{figure}

However, when $c$ becomes large, $\tau\approx\frac{1}{c}$.
Another feasible way is to empirically find a $\threshold$.
In our experiment, we assume the attacker owns another classification dataset Fashion-MNIST~\cite{XRV17}.
For each candidate model, the attacker first trains five student models on the Fashion-MNIST dataset. 
Then, for a specific $\threshold$, the attacker randomly chooses around 50\% teacher candidates from the candidate set (3/7 in our case), and launches inference attacks against the student models. 
After repeating this process 20 times, the attacker calculates the true positive rate (TPR) and false positive rate (FPR), where we assume that the student coming from a teacher candidate is a positive event.
By increasing $\threshold$ from 0 to 1, the attacker can obtain the ROC (receiver operating characteristic) curve for analysis, as presented in~\autoref{figure:threshold}. 
$\threshold=0.3$ is a good choice, as it achieves a high TPR and a low FPR. 
Finally, we set $\threshold$ as $\max(\frac{1.5}{c}, 0.3)$.

%==================================================
\subsection{Evaluation Results}
%==================================================

%==================================================
\subsubsection{Overview}
%==================================================

\begin{figure*}[!t]
\centering
\includegraphics[width=1.8\columnwidth]{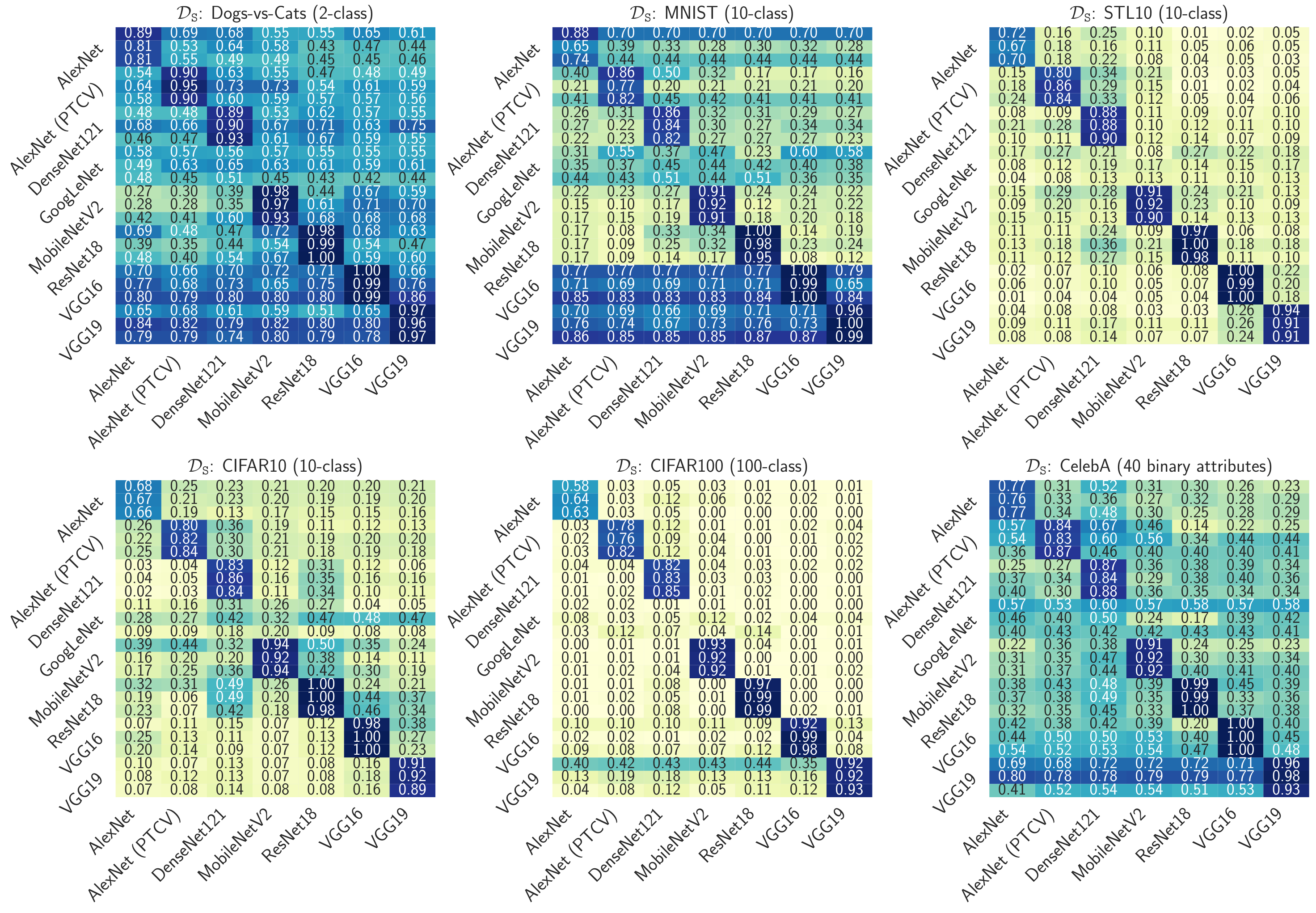}
\caption{
Teacher model fingerprinting vectors w.r.t.\ different classification tasks (100 fingerprinting pairs for each teacher model candidate, in the feature extractor setting). 
The x-axis represents the candidate teacher model, while the y-axis represents the actual teacher model. 
Each row refers to a $\fingerprintingvector$, and every three adjacent rows correspond to three different student models from the same teacher model, e.g., the first three rows of each subfigure represent three different models transferred from AlexNet. 
Note: ``GoogLeNet'' in the y-axis shows the case that the actual teacher model does not belong to the candidate set (i.e., models annotated by the x-axis). 
In this case, our attack against 13 of 18 GoogLeNet student models returns \texttt{NULL} (i.e., ``no answer'').
}
\label{fig:fixed_feature_extractor}
\end{figure*}

We exhibit our basic experiment results on transfer learning targets in~\autoref{fig:fixed_feature_extractor}.
Each row of the subfigure represents a fingerprinting vector $\fingerprintingvector$ against a specific student model, in which each column reports the matching proportion $\proportion$ on one teacher feature extractor candidate.
For all the 126 victim student models transferred from the seven teacher model candidates, the teacher model candidates with the highest $\proportion$ are consistent with the ground truths (i.e., $100\%$ inference accuracy).
Also, our attack against 13 out of 18 GoogLeNet student models and 31 out of the 36 trained-from-scratch models returns \texttt{NULL}. 
For the ease of analysis, in the remaining parts, we mainly consider the case that all the possible teacher candidates are obtained by the attacker, which is a common assumption by existing work~\cite{WYVZZ18,YLZZ19}.

Our first observation is that, in general, the more classes that the transfer learning task involves, the stronger evidence the attacker can receive to infer the teacher model.
We can see that for the CIFAR100 learning task, the highest $\proportion$ is significantly higher than other elements of $\fingerprintingvector$, which conveys strong evidence that the student model is likely to be transferred from the corresponding teacher model.
In the meanwhile, the evidence drawn from the Dogs-vs-Cats student models is not so obvious, as we can see high $\proportion$ on multiple teacher feature extractor candidates.
For instance, when sending fingerprinting pairs generated from VGG19 to the three black boxes transferred from the VGG16, the attacker has achieved $\proportion$ as 0.66, 0.76, and 0.86, respectively. 
They are relatively high compared with these on the CIFAR100 targets (0.13, 0.04, and 0.08).

Our second observation is that the student dataset possibly affects the attack performance.
For the four 10-class classification tasks, we can see that the discrimination between elements of $\fingerprintingvector$ in an attack against MNIST student models is less evident than in the other three kinds of transfer learning student models.
In fact, the attack performance is subject to the similarity between the student dataset and the probing dataset.
We will study this phenomenon in~\autoref{section:inference_resource}.

%==================================================
\subsubsection{For Unknown $k$}
\label{section:probing_features}
%==================================================

In practice, the number of components $k$ used by the teacher model is usually available when the teacher model gets released~\cite{YLZZ19}. 
But it is still worth investigating the rare case when $k$ is unknown.
In this case, we choose the following strategy to identify $k$:
for all possible values of $k$, the attacker synthesizes attack queries.
Then, the attacker launches attacks with different $k$, from the smallest value to the largest value.
For each $k$, if the predicted result is not \texttt{NULL}, we will record the current candidate model and the corresponding $k$ as the inference result.
Otherwise, the inference process will stop.

\mypara{Setup}
Basically, in our experiment, we group pre-trained layers by separating them by the pooling layers, and we call each separated part a ``block.''
In our experiment, we freeze or fine-tune blocks instead of individual layers. 
We report the number of blocks for each $\featureextractor$ in our experiment in~\autoref{table:teacher_model}.
So, in our case, we choose to identify the number of pre-trained blocks instead of the number of pre-trained layers $k$.
In our experiment, we consider teacher models removing the last pre-trained block and the last two pre-trained blocks, since in practice, deep features are wanted~\cite{YLZZ19}.
It is notable that we freeze all pre-trained components in this experiment.

\mypara{Results}
For the target models removing the last block, we achieve a 100\% (126/126) inference accuracy.
Meanwhile, for the target models removing the last two blocks, we achieve a 65.87\% (83/126) inference accuracy.

One possible illustration of this phenomenon is that the earlier layers extract more generic visual features from the input, such as edges, dots, and textures.
In contrast, the deep layers extract more abstract features (like complex patterns, objects, and even concepts), which is more specific knowledge of the teacher dataset.
That means features from deeper layers carry the information on how the teacher model refines knowledge from inputs, and therefore they are the key to extracting teacher model fingerprints.
We can think that most part of the teacher model fingerprint is removed when deep layers get discarded.

\begin{figure}[!htb]
\centering
\includegraphics[width=.83\columnwidth]{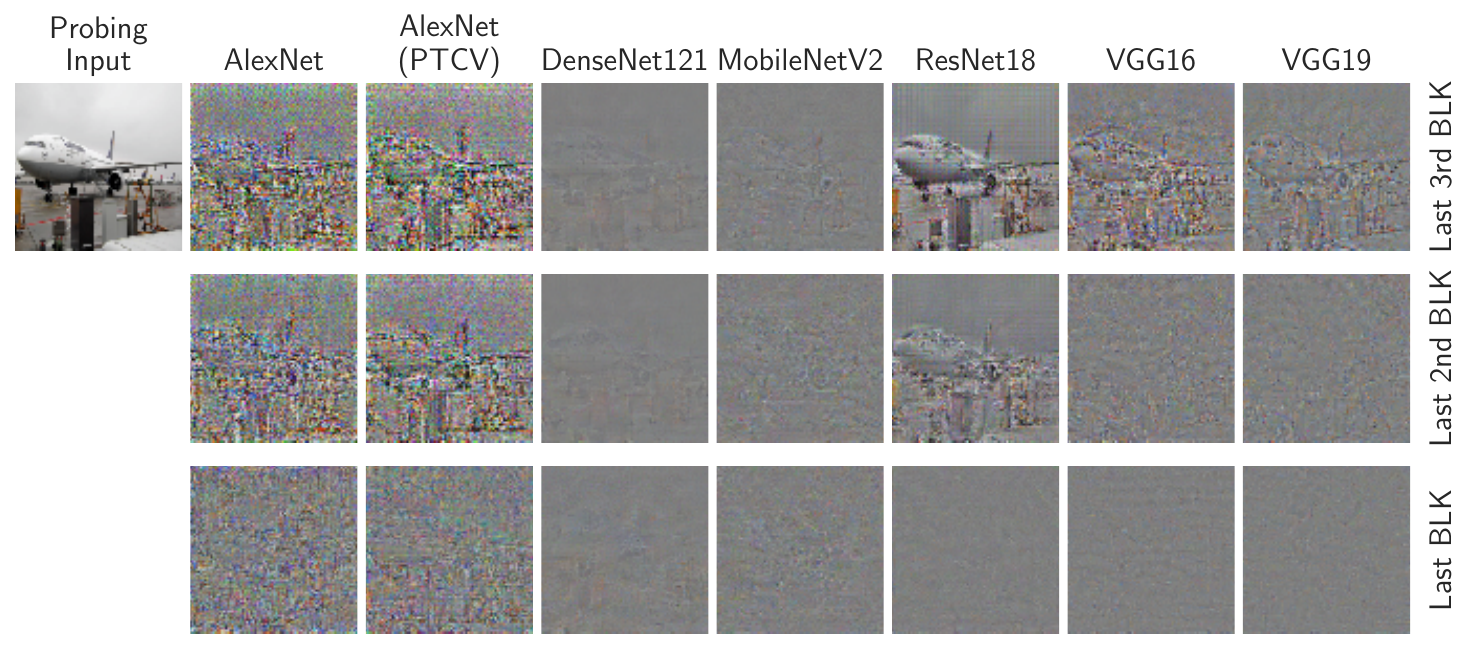}
\caption{
An example of how probing features affect synthetic inputs.
The picture at the left top corner is the original probing input.
Each row refers to synthetic inputs generated by features from different blocks, and each column refers to synthetic inputs for a specific candidate teacher model.
}
\label{fig:example_wrt_reference_features}
\end{figure}

\autoref{fig:example_wrt_reference_features} exhibits some synthetic inputs generated from different blocks.
We can observe that synthetic inputs generated from lower-level blocks tend to present more concrete contents,
i.e., the ``plane'' pattern presented in the probing input.
Meanwhile, synthetic inputs from high-level blocks look more like abstract ``noisy patterns.''
To understand this phenomenon, we need to review how features are extracted and propagated layer by layer.
On early layers, local patterns of the synthetic input are captured, while on deep layers, a global and abstract description of the input is extracted.
As a result, only if the synthetic inputs contain sufficient visual details and local patterns (edges, dots, etc.) can they activate similar low-level features with the probing inputs.
Considering an extreme case: when we generate a synthetic input with the probing input on the input layer, we tend to produce a copy of the probing input.
Consequently, they are more likely to result in matched responses on different models, because they are close to the probing inputs on too many low-level features.
It leads to more unwanted matches on victims transferred from other teacher candidates.
Hence, it is critical to extract fingerprints from deep layers.

\subsubsection{Impact of Fine-Tuning}
\label{section:fine_tuned_extractor}

Fine-tuning is a commonly adopted strategy for better model performance or faster model convergence~\cite{EMBBV09}.
Here, we explore how fine-tuning techniques affect the attack.

\mypara{Setup}
Recall that we divide pre-trained feature extractors into blocks. 
We study the following cases: frozen feature extractor, and the last one to five blocks get fine-tuned.
In general, when more blocks get fine-tuned, more disturbances are introduced to the pre-trained model.

\begin{figure}[!htb]
\centering
\begin{subfigure}{0.45\columnwidth}
\includegraphics[width=\columnwidth]{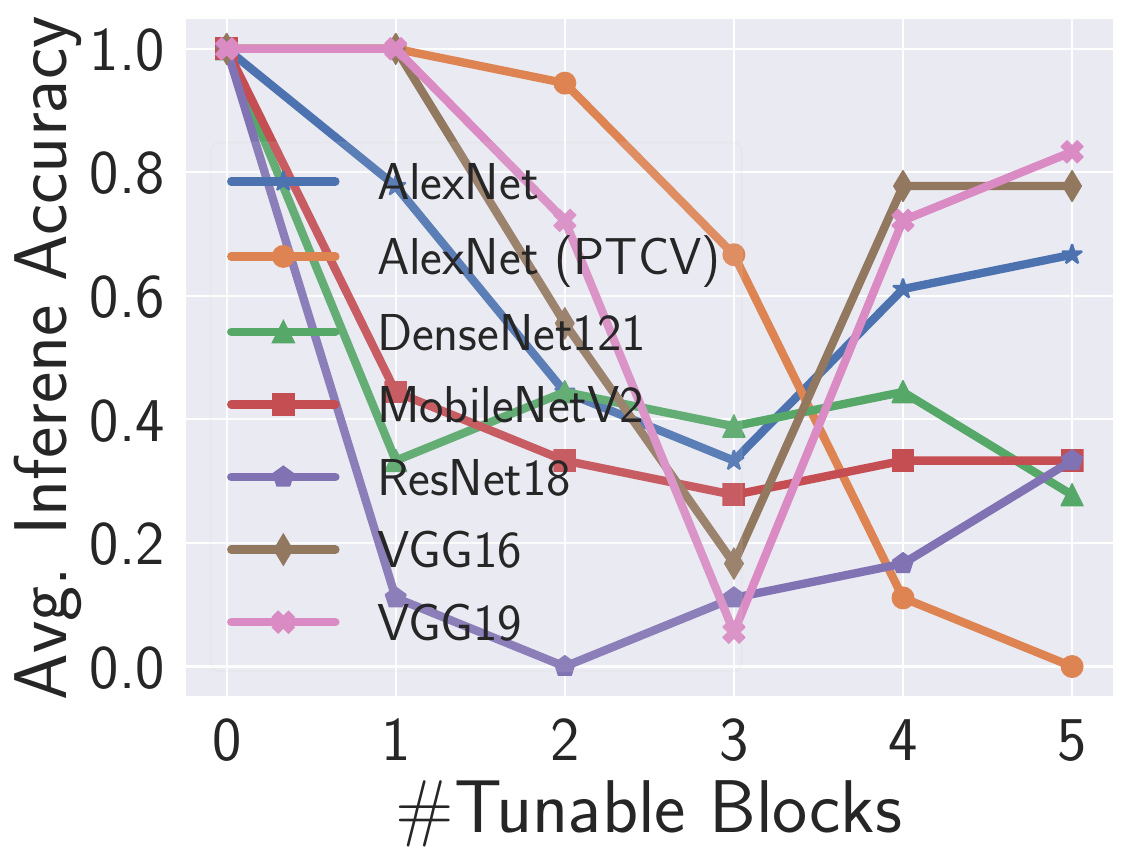}
\caption{}
\label{fig:tunable_blocks_avg_acc}
\end{subfigure}
\begin{subfigure}{0.45\columnwidth}
\includegraphics[width=\columnwidth]{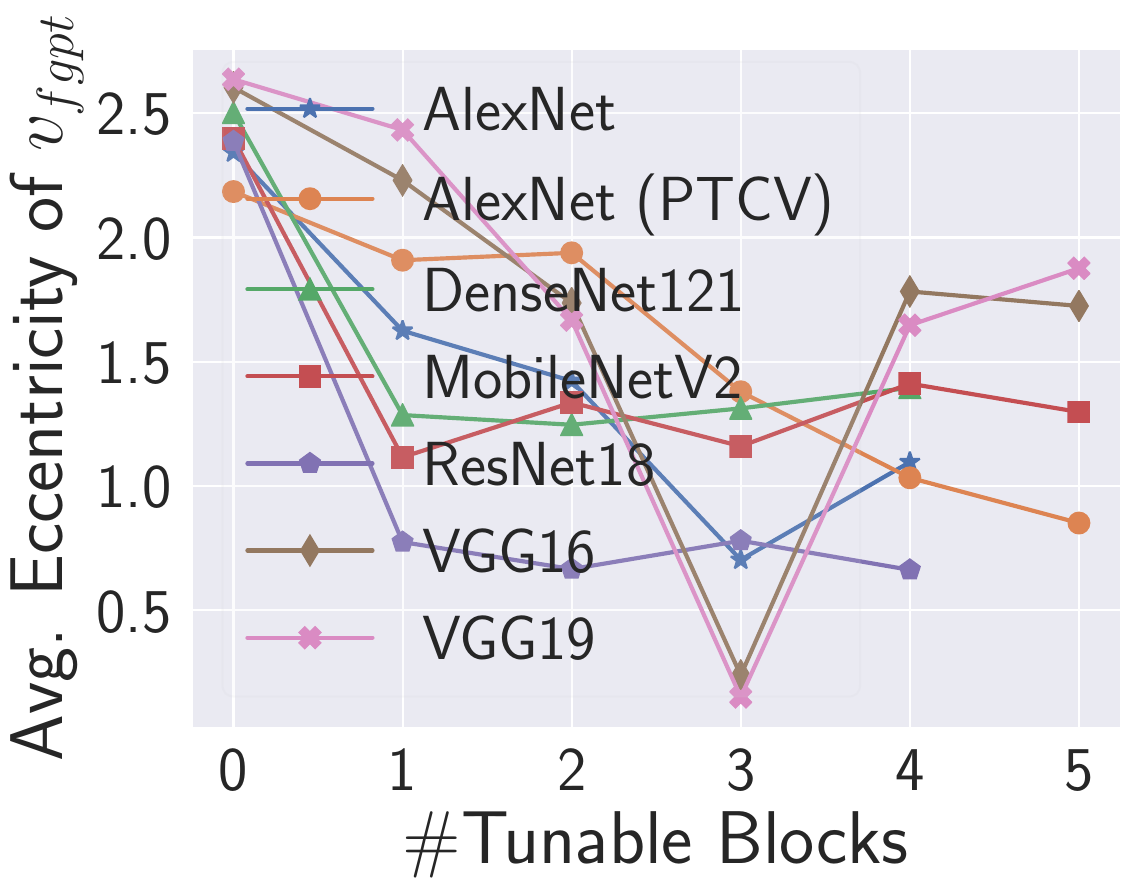}
\caption{}
\label{fig:tunable_blocks_avg_ecc}
\end{subfigure}
\caption{
Performance on fine-tuned feature extractor layers: 
(a) average inference accuracy and 
(b) average eccentricity of $\fingerprintingvector$ w.r.t.\ seven different $\featureextractor$. 
Each line corresponds to one teacher model.
When fine-turning the last four or the last five blocks, we decrease the learning rate from $10^{-3}$ to $10^{-5}$, so as to achieve a stable learning process.
Such learning rate change leads to some turning points for x=3 in the plot.
For $x$=4 and $x$=5 in the plot, though the number of fine-tuned parameters grows up, a lower learning rate reduces the absolute change of these parameters, which might lead to a higher inference accuracy.
}
\label{fig:tunable_blocks}
\end{figure}

\mypara{Results}
We evaluate the average inference accuracy for each case and plot them in~\autoref{fig:tunable_blocks_avg_acc}.
In general, the inference accuracy is likely to decrease when more parameters get fine-tuned.
We present fingerprinting vectors when one and two blocks get fine-tuned in~\autoref{fig:last_one_fine_tuned} and~\autoref{fig:last_two_fine_tuned}.
We also statistically study the eccentricity of $\fingerprintingvector$ as exhibited in~\autoref{fig:tunable_blocks_avg_ecc}. 
Our key observation is that the eccentricity of fingerprinting vectors significantly decreases when we fine-tune pre-trained layers of ResNet18, MobileNetV2, and DenseNet121,
while fine-tuning high-level pre-trained layers of other targets have less impact on the attack performance.
After checking the pre-trained model architecture, we find another potential cause of this phenomenon: ResNet18, MobileNetV2, and DenseNet121 contain batch normalization layers between convolution layers.
When fine-tuned on new datasets, the empirical mean and variance of the batch normalization layer are updated gradually. 
It may cause a significant change in the latent feature representation, breaking the basic assumption that the $\featureextractor'(\cdot)\approx\featureextractor(\cdot)$.

%==================================================
\subsubsection{Impact of Query Budget}
\label{section:query_budget}
%==================================================

\begin{figure}[!t]
\centering
\includegraphics[width=1\columnwidth]{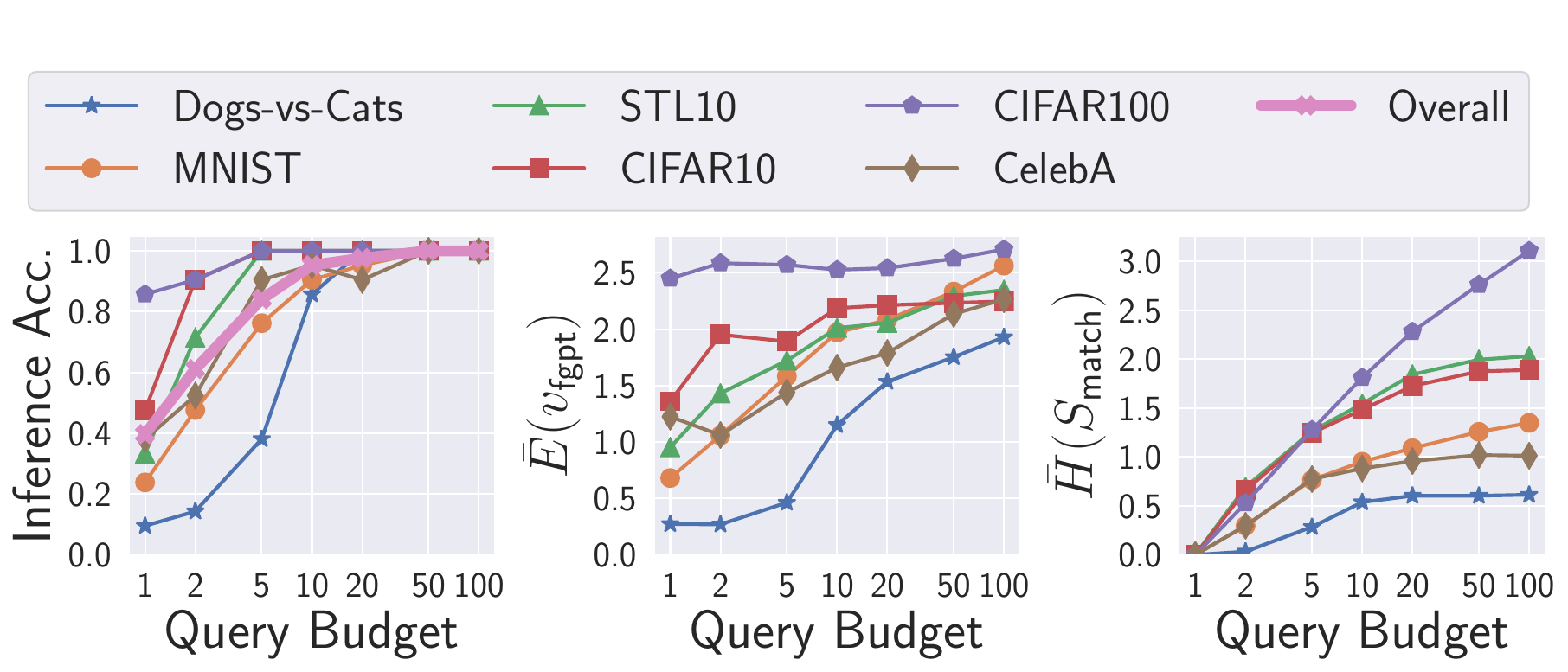}
\caption{
Performance on different query budgets. 
Each line depicts the attack performance on a specific transfer learning task.
We have achieved 100\% inference accuracy when the query budget reaches 50 (pairs for each teacher model candidate). 
On transfer learning tasks with more complex input patterns and more classes, we get higher eccentricity and entropy heuristics.
}
\label{fig:performance_wrt_budget}
\end{figure}

Then query budget is a primary concern of a query-based attack.
A low query budget will not only limit the attack cost, but also help the attacker to keep stealthy.

\mypara{Setup}
In our experiment, we examine the effect of the query budget by simulating attacks that send 1, 2, 5, 10, 20, 50, and 100 fingerprinting pairs for each teacher candidate.

\mypara{Results}
\autoref{fig:performance_wrt_budget} depicts attack performance on different query budgets, i.e., the number of fingerprinting pairs for each teacher model candidate sent to the target black box.
Therefore, the total number of queries is $2 * \text{query budget} * |\{\featureextractorNumbered{j}\}|$.
Overall, as the query budget grows, the inference accuracy increases rapidly.
When the query budget reaches 50, the total inference accuracy achieves 100\%. 
It suggests that our attack can avoid possibly high query charge caused by existing model reverse engineering attacks~\cite{TZJRR16,OASF18}.
We also see that the attack on more complex tasks (more classes and more complicated inputs) tends to achieve higher inference accuracy. 
For example, inference accuracy on CIFAR100 classifiers exceeds 80\% even the query budget is limited to 10.

To gain a deeper insight, we also examine the average eccentricity of $\fingerprintingvector$ and the average entropy of $\matchset$ with the highest $\proportion$.
We find that there possibly exist some correlations among eccentricity, entropy heuristics, and inference accuracy. 
Firstly, the attack against a more complex transfer learning task is likely to bring higher ``inference belief.''
For instance, the attack on CIFAR100 classifiers shows relatively high average eccentricity and entropy compared to other learning tasks.
Secondly, we can observe that as the query budget increases, the average eccentricity of $\fingerprintingvector$ and the average entropy of $\supportset$ also rise.
This phenomenon is consistent with our intuition: more queries are supposed to return more ``evidence'' to the attacker (higher entropy), and thus, the attacker has more confidence (higher eccentricity) to perform the inference, and they have a higher chance to hit the correct answer (higher accuracy).

%==================================================
\subsubsection{Impact of Probing Inputs}
\label{section:inference_resource}
%==================================================

One of our basic attack assumptions is that any information about the population of $\studentdataset$ is unavailable.
Previously, we use probing inputs from a public dataset, VOC-Segmentation, which is not overlapped with $\studentdataset$.
But what if when there exists a ``stronger'' attacker, who has collected probing inputs following the same distribution with $\studentdataset$; or there is a ``weaker'' attacker, who has no access to any realistic data?

\mypara{Setup}
To this end, we investigate how probing datasets affect attacks with extra evaluations on the other three probing datasets:
\setlist{nolistsep}
\begin{itemize}[noitemsep]
\item \mypara{MNIST and CelebA} They are realistic images and share the same distribution with $\studentdataset$. We randomly choose 100 probing inputs from the MNIST and CelebA testing dataset, respectively. 
In this case, the $\studentdataset$ and the input probing dataset are still not overlapped.
\item \mypara{Random Noise} This is an extreme case where no realistic images are available. We randomly sample 100 probing inputs from the normal distribution.
\end{itemize}

\mypara{Results}
Overall, using the VOC-Segmentation dataset, in general, achieves good performance, while Random Noise leads to the worst performance.
We plot the attack performance with MNIST, CelebA, and Random Noise probing data in~\autoref{fig:performance_with_MNIST},~\autoref{fig:performance_with_celeba}, and~\autoref{fig:performance_with_noise} respectively.

The major obstacle to using Random Noise data as probing inputs is the low entropy of $\matchset$.
We find that the target model often classifies Random Noise inputs to a specific label, leading to a low entropy value of $\matchset$.
Meanwhile, when we choose MNIST as the probing dataset, we have achieved high entropy values and 100\% inference accuracy (query budget is 100) on the MNIST-classification targets.
We have similar observations when using CelebA as probing inputs for CelebA-classification models.
This is expected: with sufficient probing inputs from the same distribution with $\studentdataset$, the attacker can increase the variety of output labels and increase the entropy of $\matchset$.
Consequently, they have more information to make a reliable inference.
We will further analyze how the probing inputs affect the attack in~\autoref{section:statistical_analysis}.

%================================================== 
\section{More Robust Teacher Model Fingerprinting}
\label{section:statistical_analysis}
%==================================================

The experimental results exhibited in~\autoref{section:evaluation} have shown that the na\"ive ``one-of-the-best'' strategy is effective in most cases.
However, we can also find that it is possible to obtain a high $\proportion$ by a mismatched feature extractor.
For instance, in~\autoref{fig:fixed_feature_extractor}, for MNIST classification student models, we get $\proportion=0.86$ with an AlexNet candidate against a VGG19-based student model.
It indicates there would exist a considerable number of ``false matches'' in the matching set $\matchset$.
One possible reason for the false-matching problem is that most attack queries (especially the synthetic input) belong to unrecognizable contents of the target black box.
In this scenario, the target classifies a pair of ``meaningless'' inputs into the same class occasionally, even if they produce very different features.
We can understand this phenomenon in an extreme case: supposing a black box only returns the label ``1'' to any input.
In this case, all our sent fingerprint pairs will activate seemingly perfect matched responses, and we will surely get $\proportion=1$ for each candidate teacher feature extractor.
But it is obvious that we cannot find out the actual teacher model as we are unable to affect the black-box's behavior.

We design inference strategies leveraging statistical testing methods to alleviate the impact of false-matching problems. 
Here we firstly give one definition for further discussion.

\begin{definition}[Supporting Set]
\label{definition:supporting_set}
After removing the elements with the maximum occurrence from $\matchset$, the remaining elements compose the supporting set $\supportset=\{\modeloutput| \forall \modeloutput \in \matchset, \modeloutput \neq \argmax_{i}|\{\forall\modeloutput_i\in\matchset\}|\}$, which will be leveraged as the evidence to infer the original model.
\end{definition}

\begin{figure}
\centering
\includegraphics[width=0.65\columnwidth]{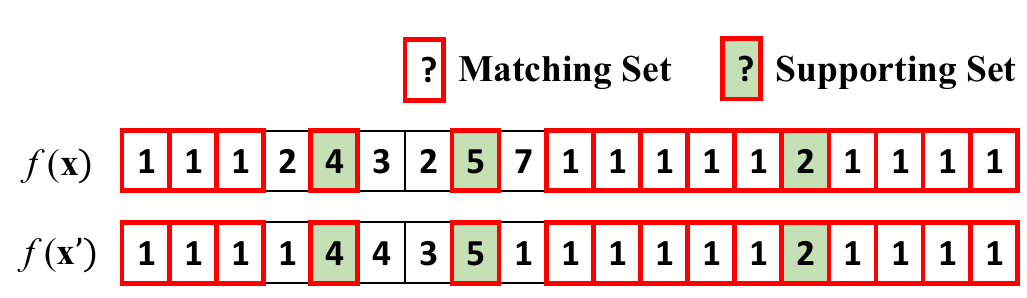}
\caption{
Illustration of the supporting set. The supporting set is the collection of the remaining elements after removing the elements with the maximum occurrence from $\matchset$.
}
\label{fig:illustration_supporting_set}
\end{figure}

The relationship between the fingerprinting attack responses, matching set, and supporting set is explained by~\autoref{fig:illustration_supporting_set}.

Then we consider the extreme case that the student model $\studentmodel$ has zero knowledge about the input. 
It randomly divides the input space into $c$ classes, which we call a \textit{random classifier}.\footnote{We emphasize that the word ``random'' here refers to that the parameters of $\studentmodel$ are randomly configured with respect to $\studentdataset$, rather than it outputs label randomly given a specific input.}
So how much is the possibility of getting a match on a fingerprinting pair?
We formalize the following statistical hypothesis testing: given a candidate teacher model $\teachermodel$, a student model $\studentmodel$, a set of fingerprinting pairs $\{\pair{\modelinput_i, \modelinput_i'}\}$ and the corresponding responses $\{\pair{\modeloutput_i, \modeloutput_i'}\}$, we make the following hypothesis
\[
\begin{aligned}
    &H_0: \textit{$\studentmodel$ is a random classifier.}\\
    &H_1: \textit{$\studentmodel$ is not a random classifier.}
\end{aligned}
\]

We can give a sufficient condition to reject the null hypothesis $H_0$ give a specific significance value $\alpha$.
\begin{theorem}
Given a $c$-class classification student model $\studentmodel$ in top-1 label exposure, a fingerprinting statistical hypothesis ($H_0$, $H_1$, as well as the significance value $\alpha$), the size of the supporting set sufficient to reject the null hypothesis $H_0$ is
\begin{equation}
\ceil*{\log_{2}\frac{1}{\alpha}}.
\label{equ:hypothesis_testing}
\end{equation}
\end{theorem}

\proof See~\autoref{section:proof1}.

\newcolumntype{s}{>{\columncolor[HTML]{ECF0F1}} c}
\begin{table*}
\scriptsize
\centering
\begin{tabular}{c|csc|csc|csc|csc}
\toprule
\multirow{3}{*}{\makecell{\textbf{Query}\\ \textbf{Budget}}} 
& \multicolumn{3}{c}{\textbf{probing: VOCSegmentation}} 
& \multicolumn{3}{c}{\textbf{probing: MNIST}} 
& \multicolumn{3}{c}{\textbf{probing: CelebA}} 
& \multicolumn{3}{c}{\textbf{probing: Random Noise}} \\ 
\cline{2-13} 
& \multicolumn{2}{c}{\textbf{inference acc.}} 
& \multirow{2}{*}{$\frac{\textbf{\#robust}}{\textbf{\#original}}$} 
& \multicolumn{2}{c}{\textbf{inference acc.}} 
& \multirow{2}{*}{$\frac{\textbf{\#robust}}{\textbf{\#original}}$} 
& \multicolumn{2}{c}{\textbf{inference acc.}} 
& \multirow{2}{*}{$\frac{\textbf{\#robust}}{\textbf{\#original}}$} 
& \multicolumn{2}{c}{\textbf{inference acc.}} 
& \multirow{2}{*}{$\frac{\textbf{\#robust}}{\textbf{\#original}}$} \\ 
\cline{2-3} \cline{5-6} \cline{8-9} \cline{11-12}
& \textbf{original} & \textbf{robust} & 
& \textbf{original} & \textbf{robust} & 
& \textbf{original} & \textbf{robust} & 
& \textbf{original} & \textbf{robust} & \\ 
\midrule
1 
& \Gape[0pt][2pt]{\makecell{39.68\%\\(50/126)}} & {-- (0/0)} & 0 (0/126) 
& \Gape[0pt][2pt]{\makecell{42.06\%\\(53/126)}} & {-- (0/0)} & 0 (0/126)
& \Gape[0pt][2pt]{\makecell{45.24\%\\(57/126)}} & {-- (0/0)} & 0 (0/126)
& \Gape[0pt][2pt]{\makecell{19.84\%\\(25/126)}} & {-- (0/0)} & -- (0/126) \\
2 
& \Gape[0pt][2pt]{\makecell{61.11\%\\(77/126)}} & {-- (0/0)} & 0 (0/126) 
& \Gape[0pt][2pt]{\makecell{57.94\%\\(73/126)}} & {-- (0/0)} & 0 (0/126) 
& \Gape[0pt][2pt]{\makecell{57.94\%\\(73/126)}} & -- (0/0) & 0 (0/126) 
& \Gape[0pt][2pt]{\makecell{29.37\%\\(37/126)}} & -- (0/0)& -- (0/126) \\
5 
& \Gape[0pt][2pt]{\makecell{84.13\%\\(106/126)}} & -- (0/0)& 0 (0/126) 
& \Gape[0pt][2pt]{\makecell{69.84\%\\(88/126)}} & -- (0/0) & 0 (0/126) 
& \Gape[0pt][2pt]{\makecell{80.95\%\\(102/126)}} & -- (0/0) & 0 (0/126)
& \Gape[0pt][2pt]{\makecell{42.06\%\\(53/126)}} & -- (0/0)& -- (0/126) \\
10 
& \Gape[0pt][2pt]{\makecell{95.24\%\\(120/126)}} & \Gape[0pt][2pt]{\makecell{100.00\%\\(32/32)}} & \Gape[0pt][2pt]{\makecell{25.40\%\\(32/126)}} 
& \Gape[0pt][2pt]{\makecell{80.95\%\\(102/126)}} & {\Gape[0pt][2pt]{\makecell{100.00\%\\(19/19)}}} & \Gape[0pt][2pt]{\makecell{15.08\%\\(19/126)}} 
& \Gape[0pt][2pt]{\makecell{89.68\%\\(113/126)}} & \Gape[0pt][2pt]{\makecell{100.00\%\\(3/3)}} & \Gape[0pt][2pt]{\makecell{2.38\%\\(3/126)}} 
& \Gape[0pt][2pt]{\makecell{50.79\%\\(64/126)}} & -- (0/0)& -- (0/126) \\
20 
& \Gape[0pt][2pt]{\makecell{97.62\%\\(123/126)}} & \Gape[0pt][2pt]{\makecell{100.00\%\\(97/97)}} & \Gape[0pt][2pt]{\makecell{76.98\%\\(97/126)}} 
& \Gape[0pt][2pt]{\makecell{(84.92\%\\(107/126)}} & {\Gape[0pt][2pt]{\makecell{100.00\%\\(52/52)}}} & \Gape[0pt][2pt]{\makecell{41.27\%\\(52/126)}} 
& \Gape[0pt][2pt]{\makecell{96.83\%\\(122/126)}} & {\Gape[0pt][2pt]{\makecell{100.00\%\\(87/87)}}} & \Gape[0pt][2pt]{\makecell{69.05\%\\(87/126)}} 
& \Gape[0pt][2pt]{\makecell{57.14\%\\(72/126)}} & {\Gape[0pt][2pt]{\makecell{100.00\%\\(16/16)}}} & \Gape[0pt][2pt]{\makecell{12.70\%\\(16/126)}} \\
50 
& \Gape[0pt][2pt]{\makecell{100.00\%\\(126/126)}} & {\Gape[0pt][2pt]{\makecell{100.00\%\\(125/125)}}} & \Gape[0pt][2pt]{\makecell{99.21\%\\(125/126)}} 
& \Gape[0pt][2pt]{\makecell{90.48\%\\(114/126)}} &{\Gape[0pt][2pt]{\makecell{100.00\%\\(96/96)}}} & \Gape[0pt][2pt]{\makecell{76.19\%\\(96/126)}} 
& \Gape[0pt][2pt]{\makecell{99.21\%\\(125/126)}} & {\Gape[0pt][2pt]{\makecell{100.00\%\\(117/117)}}} & \Gape[0pt][2pt]{\makecell{92.86\%\\(117/126)}} 
& \Gape[0pt][2pt]{\makecell{62.70\%\\(79/126)}} & {\Gape[0pt][2pt]{\makecell{100.00\%\\(36/36)}}} & \Gape[0pt][2pt]{\makecell{28.57\%\\(36/126)}}\\
100 
& \Gape[0pt][2pt]{\makecell{100.00\%\\(126/126)}} & {\Gape[0pt][2pt]{\makecell{100.00\%\\(126/126)}}} & \Gape[0pt][2pt]{\makecell{100.00\%\\(126/126)}} 
& \Gape[0pt][2pt]{\makecell{96.03\%\\(121/126)}} & {\Gape[0pt][2pt]{\makecell{100.00\%\\(114/114)}}} & \Gape[0pt][2pt]{\makecell{90.48\%\\(114/126)}} 
& \Gape[0pt][2pt]{\makecell{100.00\%\\(126/126)}}  & \Gape[0pt][2pt]{\makecell{100.00\%\\(122/122)}} & \Gape[0pt][2pt]{\makecell{96.83\%\\(122/126)}} 
& \Gape[0pt][2pt]{\makecell{65.08\%\\(82/126)}} & {\Gape[0pt][2pt]{\makecell{100.00\%\\(41/41)}}} & \Gape[0pt][2pt]{\makecell{32.54\%\\(41/126)}} \\
\bottomrule 
\end{tabular}
\caption{
Comparison between the original and robust version of the teacher model fingerprinting attack.
We have achieved 100\% inference accuracy on the remaining inferences after the hypothesis testing ($\alpha=0.01$, $\ceil*{\log_{2}\frac{1}{\alpha}}=7$).
}
\label{table:safer_inference}
\end{table*}

\begin{figure*}[!htb]
\centering
\includegraphics[width=1.6\columnwidth]{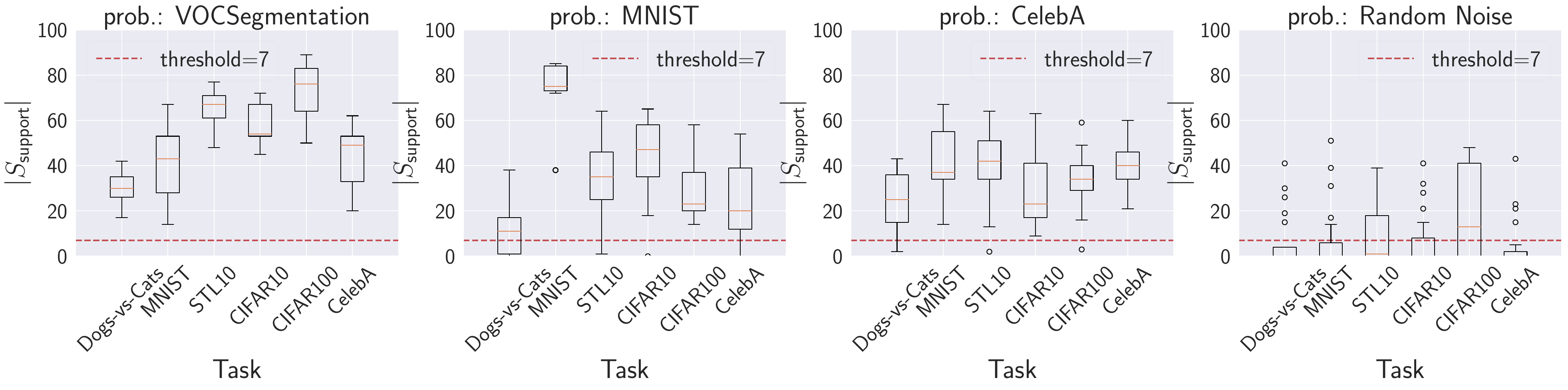}
\caption{
Supporting set size w.r.t.\ different learning tasks and different probing inputs (query budget=100, $\alpha=0.01$, the hypothesis testing threshold $\ceil*{\log_{2}\frac{1}{\alpha}}=7$). 
If the probing inputs follow a similar distribution with $\studentdataset$, $|\supportset|$ is relatively high.
}
\label{fig:supporting_set_size}
\end{figure*}

Only when $H_0$ is rejected can we make a more robust inference. 
That is, the attacker has high confidence (1-$\alpha$) to avoid a wrong inference against a random classifier.

Here we provide a more robust version of the teacher model fingerprinting attack: (1) generating fingerprinting pairs $\rightarrow$ (2) selecting the best candidate $\rightarrow$ (3) if $H_0$ is rejected, accepting the inference result; otherwise, sending more queries and repeating, until reaching the maximum query budget.  

\mypara{Evalution} 
We repeat several empirical studies on our robust attack method.
We set the significance value $\alpha$ as 0.01, and thus, the testing hypothesis threshold $\ceil*{\log_{2}\frac{1}{\alpha}}=7$.
\autoref{table:safer_inference} exhibits the robust version of inference results for~\autoref{section:inference_resource}.
Overall, we have achieved 100\% inference accuracy on the robust inferences that reject $H_0$ with $\alpha=0.01$, even without realistic datasets (Random Noise as the probing dataset).
We also examine the size of the supporting set on our inference results in~\autoref{section:inference_resource} to investigate how probing inputs and transfer learning tasks affect the attack performance.

\mypara{Impact of probing inputs}
The probing input distribution indeed impacts the attack performance.
From~\autoref{fig:supporting_set_size}, we can see that, with realistic probing inputs, the proportions of robust inferences are obviously higher than that with Random Noise. 
This is consistent with our observation in~\autoref{section:inference_resource}.
Moreover, a dataset that shares a similar distribution with $\studentdataset$ is more helpful in reducing the query budget, because it can increase the variance of the black-box responses.
For example, for MNIST classifiers, the supporting set size $|\supportset|$ significantly rises when we use MNIST as the probing dataset.

\mypara{Impact of transfer learning tasks}
\autoref{fig:supporting_set_size} shows that more complex learning tasks (i.e., those have more complex inputs and more classes) tend to produce a larger $|\supportset|$.
For instance, when attacking CIFAR100 classifiers, most inferences successfully reject $H_0$ as their $|\supportset|$ are over the threshold. 
One reason is that a complex learning task is likely to give a higher variance of the responses.
This result is consistent with our observation in~\autoref{section:query_budget}.
To help readers understand this phenomenon, we can estimate the minimal size of $\supportset$.
Since~\autoref{equ:hypothesis_testing} points out that the minimal $|\supportset|$ should be $\ceil*{\log_{2}\frac{1}{\alpha}}$, for a $c$-class classifier, $|\matchset \setminus \supportset|$ should be no less than $\ceil*{\frac{\ceil*{\log_{2}\frac{1}{\alpha}}}{c-1}}$.
Therefore, we have 
\begin{equation}
    |\matchset| \geq\ceil*{\log_{2}\frac{1}{\alpha}} + \ceil*{\frac{\ceil*{\log_{2}\frac{1}{\alpha}}}{c-1}}.
    \label{equ:support_set_minimal_size}
\end{equation}
So, according to ~\autoref{equ:support_set_minimal_size}, we can find that when $c$ increases, we will get a lower bound of the size of the supporting set. 
It should be pointed out that this is just a sufficient condition to reject $H_0$ for the robust version of our attack.

%================================================== 
\section{Applications of the Fingerprinting Attack}
%==================================================

In this section, we demonstrate how our attack helps to mount model stealing and identify vulnerable teacher models.

%==================================================
\subsection{Practical Model Stealing}
%==================================================

\begin{figure*}[!htb]
\centering
\includegraphics[width=1.9\columnwidth]{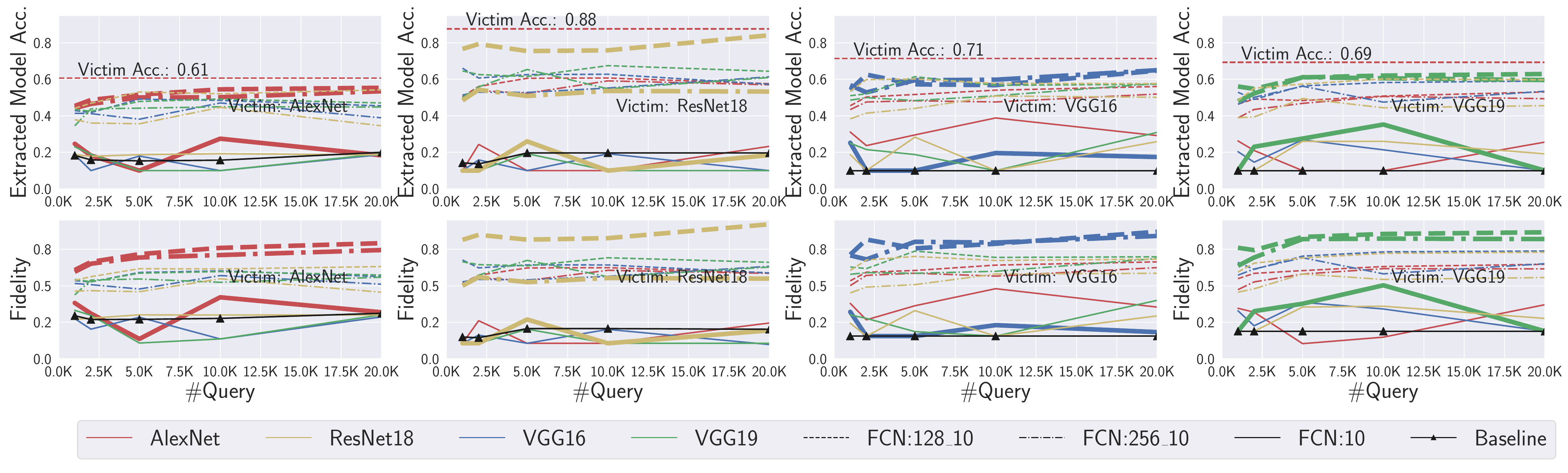}
\caption{
Model stealing against transfer learning models.
The top row reports the accuracy metric, while the bottom row reports the fidelity metric of extracted models, respectively.  
For an extracted model transferred from the same teacher model with the target (i.e., inferred by a successful teacher model fingerprinting attack; marked with thicker lines), it has finally achieved higher accuracy and fidelity with sufficient queries (\#Query $\geq$ 10,000).
}
\label{fig:model_stealing}
\end{figure*}

\mypara{Setup}
The basic model stealing strategy is to firstly identify the teacher feature extractor of the victim, and then build up a surrogate student model based on it.
Next, the attacker sends queries to the black-box target and collects the top-1 label responses.
Finally, the attacker trains the surrogate student model with the attack queries and corresponding responses.
In our experiment, we choose four CIFAR-10 classifier victims, which are transferred from AlexNet, ResNet18, VGG16, and VGG19, respectively.
The victim models own the same fully connected layer architecture: \fullyconnected{128} $\rightarrow$ \batchnorm $\rightarrow$ \softmax{10}.
We choose training images from CIFAR100 as the attack queries (non-overlapping with the target's $\studentdataset$, but sharing a similar distribution).
We evaluate the model stealing performance with four pre-trained models: AlexNet, ResNet18, VGG16, and VGG19.
To examine the affect by surrogate model architectures, for each pre-trained model we build three surrogate model with three different fully connected layer architectures:
\softmax{10}, \fullyconnected{128} $\rightarrow$ \batchnorm $\rightarrow$ \softmax{10}, and 
\fullyconnected{256} $\rightarrow$ \batchnorm $\rightarrow$ \softmax{10}.
Additionally, for each victim model, we train a surrogate model with the same model architecture but from scratch as the baseline.
Each surrogate student model is trained for 20 epochs with batch size as 48.

\mypara{Evaluation metric}
We choose two evaluation metrics widely adopted by prior studies on model stealing attacks~\cite{TZJRR16, JCBKP20}:
\setlist{nolistsep}
\begin{itemize}[noitemsep]
\item \textbf{Accuracy}: accuracy of the extracted model on inputs.
It measures how well the extracted model makes correct predictions on incoming inputs.
This metric usually lies in the context where an attacker aims to steal the strong prediction ``power'' of the victim black-box ML service with only limited efforts.
\item \textbf{Fidelity}: agreement between the target and the extracted model on inputs~\cite{TZJRR16}. 
It measures the behavior similarity between the victim model and the extracted model. 
This metric usually lies in the context where an attacker aims to faithfully replicate the ``functionality'' of the black-box victim, in particular, these potential vulnerabilities.  
\end{itemize}
We evaluate the extracted models on CIFAR10 testing set non-overlapping with the victim training set or the query set.

\mypara{Results}
\autoref{fig:model_stealing} shows the model stealing attack performance with different pre-trained models.
We highlight model stealing results after successful teacher model fingerprinting attacks, i.e., the extracted model and the victim share the same teacher model, with thicker lines.
We can observe that for model stealing after correctly identifying the teacher model, the fidelity and accuracy are obviously higher than the baseline. 
Moreover, we can find that even if the surrogate model has a different architecture of the fully connected layers, we may also achieve a relatively good model stealing.

%==================================================
\subsection{Identifying Vulnerable Teacher Models}
%==================================================

As vulnerable teacher models pose severe threats to downstream applications~\cite{YLZZ19,ZSLBY20}, there is a surging need for solutions to detect the existence of vulnerable teacher models.
We investigate the feasibility of using our attack to detect if a student model comes from a known vulnerable teacher model.

\mypara{Setup}
We adopt the method proposed by Yao et al.~\cite{YLZZ19} to construct teacher models with latent backdoors.
We choose MNIST, CIFAR10, and CIFAR100 as the dataset in our experiment. 
Then, we build the backdoored teacher model for each dataset from pre-trained AlexNet, AlexNet (PTCV), and ResNet18, respectively. 
Following a similar setup in~\cite{YLZZ19}, we first use data from class 0-4 to train the original teacher model, and use data from class 5-9 to train the student model.
We choose class 5 as the target class in all our backdoor attacks.
For each backdoored teacher model, we train three student models with different fully connected layer architectures.
As a result, there are nine backdoored student models for each dataset, nine clean student models, and six candidate models owned by the attacker.
We provide more details of the vulnerable teacher models in~\autoref{appendix:latent_backdoor}.

\mypara{Results}
The inference accuracy of backdoored and clean teacher models is 19/27 and 15/27, respectively.
In the given case, the most challenging task is to discriminate the clean teacher model from the backdoored teacher model, as the backdoored teacher models are fine-tuned from the clean teacher models.
Nevertheless, our results still show it is possible to use our attack to identify vulnerable teacher models.

%================================================== 
\section{Discussions}
%==================================================

%==================================================
\subsection{Possible Countermeasures}
%==================================================

\mypara{Input distortion}
One potential solution is to distort inputs by inserting small random noise or performing image transformations like image cropping and resizing.
Hopefully, it would only slightly affect model performance on realistic inputs but significantly reduce the inference attack accuracy, since the latent feature is sensitive to the optimized ``noise'' pattern in a synthetic input.
One primary advantage of this strategy is that the input distortion processor can be directly plugged between the raw system input and the model input, without modifying the student models.
Also, the input distortion is lightweight and easy to implement, as most image processing or deep learning libraries support various transformation operations.

\mypara{Injecting neuron distances~\cite{WYVZZ18}}
Wang et al. propose a defense method, called \textit{injecting neuron distances},  to deviate the student model's feature map from the teacher model~\cite{WYVZZ18}.
The key is to retrain all student layers on the student dataset, with the optimization goal to minimize the cross-entropy loss while ensuring the dissimilarity between the student's and the teacher's intermediate representations is above a threshold.
In this case, the intuition behind our attack, i.e., that the teacher model and student model share a similar feature map, does not hold.
Despite the advantage of effectively disturbing the student model, \textit{injecting neuron distances} brings additional costs to update the whole student model. 

\mypara{Evaluation}
We investigate the feasibility of the two countermeasures on the CIFAR10 dataset.
For \emph{input distortion}, we randomly crop the input with 0.85 areas preserved, and then resize it to the size of the model's input layer.
For \emph{injecting neuron distances}, we slightly modify the implementation by~\cite{WYVZZ18}: instead of requiring a threshold, we directly introduce a negative neuron distance term into our learning objective: $\min~\texttt{CrossEntropy}-\lambda \cdot \texttt{NeuronDistance}$.
In our experiment, we set $\lambda=10^{-6}$.

To evade the defense, one attack strategy is to generate attack queries from shallower layers, so as to reduce the impact of parameter changes. 
We investigate the robustness of countermeasures against attack queries generated from different positions. 
We report the results in~\autoref{table:defense}.
We can observe that the two countermeasures can effectively defend our attack.
Another possible adaptive attack is to perform the optimization~\autoref{equ:query_generation} simultaneously across different pre-trained layers.
However, it would intensively increase the overhead.
We will explore adaptive attacks in the future.

\begin{table}[t]
\scriptsize
\centering
\begin{tabular}{cccc}
\toprule
\multirow{2}{*}{Defense} & \multicolumn{3}{c}{Attack Accuracy (Queries generated from)} \\
\cline{2-4} & Last BLK & Last 2nd BLK & Last 3rd BLK \\
\midrule
Input Perturbation & 13/21 & 11/21 & 10/21 \\
Injecting Neuron Distances & 7/21 & 5/21 & 5/21 \\
Without Defense & 21/21 & 20/21 & 15/21 \\ 
\bottomrule
\end{tabular}
\caption{
Performance of the two countermeasures.
The average testing accuracy of models that are unprotected, protected by \emph{input perturbation}, and protected by \emph{injecting neuron distances} are 77.5\%, 72.3\%, and 79.1\%, respectively.
Note that since the \emph{injecting neuron distances} trains all the model parameters, which may work like fine-tuning.}
\label{table:defense}
\end{table}

%==================================================
\subsection{Limitations and Future Work}
%==================================================

\mypara{Language model}
In this paper, our empirical studies focus on computer vision tasks.
Recently, the transfer learning technique plays a vital role in a more broad range of fields.
For instance, famous pre-trained language transformers, like BERT and GPT-3, are widely adopted into various downstream applications, such as search ranking~\cite{GDC20} and medical text mining~\cite{LYKKKSK20}. 
We think it is possible to infer pre-trained language teacher models with the same strategy proposed in this paper.
We plan to extend this work into sequential language models, and devise new fingerprinting attacks. 

\mypara{Advanced adversarial attacks}
Our results show that we can lower the attack bound of model stealing attacks based on a successful teacher model fingerprinting attack.
We believe our proposed attack can assist other advanced adversarial attacks, e.g., black-box adversarial example attack~\cite{WYVZZ18} and membership inference attack~\cite{SZHBFB19}.
We will explore more adversarial attacks preceded by our attack in real-world settings.

\mypara{Fingerprint erase}
Our empirical study indicates that the teacher model fingerprint may still exist even after fine-tuning.
In the feature, we plan to study which factors may impact the model fingerprint, and explore ways to erase teacher model fingerprints from student models as fundamental defenses. 

%================================================== 
\section{Related Work}
%==================================================

\mypara{Model reverse engineering}
Model reverse engineering aims to infer model parameters, structures, or other model-related information such as hyper-parameters, by just inspecting model responses to a specific set of attack queries.

\begin{figure}
\centering
\includegraphics[width=.22\textwidth]{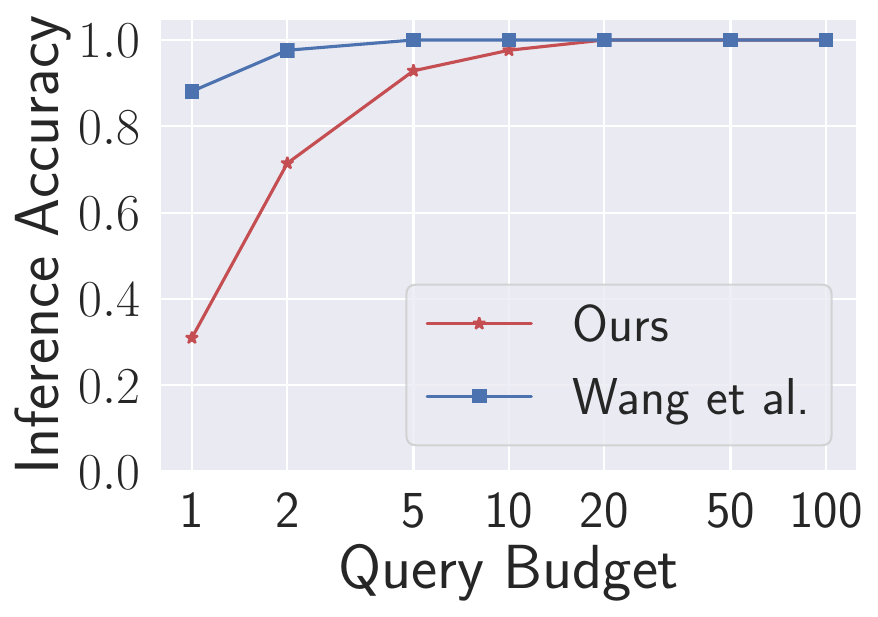}
\caption{
Comparison with the attack proposed by Wang et al.~\cite{WYVZZ18}.
To simulate the setup in~\cite{WYVZZ18}: we assume the classification scores are available; each student model is composed of a fixed feature extractor and a classification layer.
There are 42 student models in total (7 teacher models * 6 datasets).}
\label{figure:comparison}
\end{figure}

One line is to reproduce mimic models in the parameter level, i.e., they try to recover the exact model parameters.
Besides, Carlini et al.~\cite{CJM20} propose a cryptanalytic extraction approach by sending queries at critical points. 
Another line tries to duplicate target models at the functionality level.
Recently, Orekondy et al.~\cite{OSF19} propose more advanced attack methods to steal model parameters.
Jagielsk et al.~\cite{JCBKP20} discuss the inherent limitations of the learning-based model stealing strategy, and develop a more practical functionality-equivalent stealing attack. 
For transfer learning, Wang et al.~\cite{WYVZZ18} propose a teacher model fingerprinting strategy.
Its main idea is to craft a fingerprinting image to produce a nearly all-zero latent vector on the victim, when it is derived from the candidate teacher model.
Yet, it requires access to the raw confidence values.
In contrast, our attack works well when only top-1 labels are available.
\autoref{figure:comparison} compares our attack with the attack proposed by Wang et al.~\cite{WYVZZ18}.
It can be seen that our attack can achieve competitive inference accuracy when the query budget exceeds 10.
We also propose a robust fingerprinting attack to overcome the false-matching problem.
Besides, there are also efforts in inferring structures~\cite{OASF18}, hyper-parameters~\cite{WG18}, or other model properties~\cite{GWYGB18}. 

The above studies reveal that there is no such an exact line separating the white-box and black-box models.
Nonetheless, they need numerous queries and complex computations to change a black-box model into a white-box one, especially for complicated models
like VGG16.
When it comes to transfer learning, we believe our method has the advantage to quickly and efficiently reverse engineer most parts of the black-box victim---the complex transferred components, which is hard to achieve by directly combining the prior arts.

\mypara{Adversarial attacks in the transfer learning setup}
Recent studies reveal that the transfer learning technique may expose ML models to various threats.
Zhang et al.~\cite{ZSLBY20} show that adversarial examples become more transferable on a model trained by fine-tuning. 
Wang et al.~\cite{WYVZZ18} propose an adversarial example attack by crafting malicious perturbations to activate mimic latent representations to be classified to the target class.
Besides, Yao et al.~\cite{YLZZ19} present a latent backdoor attack by releasing a malicious teacher model, with which the carefully mined backdoor will infect the student model. 
In addition to the security side, transfer learning also suffers from privacy threats.
For example, Zou et al.~\cite{ZZBZ20} reveal that the public knowledge would raise the membership leakage risk in the transfer learning practice.
There also exist a broad range of other adversarial attacks against ML~\cite{FJR15, SRS17, SSSS17, SZHBFB19, NSH19, SBBFZ20, YYZTHJ20, CYZF20, ZJWG21, HJBGZ21, CSBMSWZ21}, which are out of the scope of this paper.

%================================================== 
\section{Conclusions}
%================================================== 

In this paper, we take the first step to investigate the teacher model exposure threat in the transfer learning context.
In particular, we propose a novel teacher model fingerprinting attack that can efficiently trace back the origin of a student model.
Extensive experiment results demonstrate that our attack can accurately identify the teacher model even in restrictive attack scenarios, such as top-1 label exposure and no realistic data available. 
Besides, we propose a robust attack to eliminate false positive inference results.
We also show that our attack can facilitate advanced attacks or help model forensics.
Our findings highlight the urgent need for new model confidentiality protection measures specified for transfer learning.

%================================================== 
\section*{Acknowledgments}
%================================================== 

This work is supported by the National Key Research and Development Program of China (2020AAA0107702), National Natural Science Foundation of China (U21B2018, 62161160337, 62132011), Shaanxi Province Key Industry Innovation Program (2021ZDLGY01-02), the Research Grants Council of Hong Kong under Grants N\_CityU139/21, R6021-20F, R1012-21, and the Helmholtz Association within the project ``Trustworthy Federated Data Analytics'' (TFDA) (funding number ZT-I-OO1 4).
Chao Shen, Cong Wang, and Yang Zhang are the corresponding authors.

%================================================== 
\bibliographystyle{plain}
\bibliography{normal_generated_py3}
%================================================== 

%================================================== 
\appendix
%================================================== 

%================================================== 
\section*{Appendix}
%================================================== 

%================================================== 
\section{Proof of Theorem 1}
\label{section:proof1}
%================================================== 

\begin{proof}
For a $c$-class classifier $\studentmodel$ whose response is only the top-1 label (i.e. $\modeloutput=\argmax_i \studentmodel(\modelinput)_i$), define

\begin{equation}
    p_k=\int_{\modelinput\in\inputspace}\prob{\argmax_i \studentmodel(\modelinput)_i==k}, \forall k \in [1, 2, \cdots, c],
\end{equation}
where $\inputspace$ refers to the input space.

Without loss of generality, we assume that $1 \geq p_1 \geq p_2 \geq \cdots \geq p_c \geq 0$.\footnote{In most cases, $p_c>0$. However, here we consider a more general case, where the black box's output space may not cover all the $c$ classes.}
Also noticing that $\sum_{i=1}^{c}p_i=1$, we have

\begin{equation}
\begin{aligned}
    1=\sum_{i=1}^{c}p_i \geq \sum_{i=1}^{k}p_i \geq kp_k 
    \Rightarrow p_k \leq \frac{1}{k}.
\end{aligned}
\end{equation}

Suppose the attacker generates $N$ fingerprinting pairs $\{\pair{\modelinput_i,\modelinput_i'}\}$ with a set of probing input $\{\modelinput_i\}$, and receives $N$ pairs of responses $\{\pair{\modeloutput_i,\modeloutput_i'}\}$, among which $M$ pairs satisfy $\modeloutput==\modeloutput'$. 
For simplicity, we further suppose the first $M$ fingerprinting pairs produce the matched responses (i.e., $\modeloutput_i==\modeloutput_i'$ when $ 1\leq i\leq M$, and $\modeloutput_j\neq\modeloutput_j'$ when $M+1 \leq j \leq N$).
If the victim is a random classifier, we have 

\begin{equation}
\begin{aligned}
    \prob{\{\modeloutput_i'\}|\{\pair{\modelinput_i, \modelinput_i'}\}, \{\modeloutput_i\}, \studentmodel}
    &=\prod_{i=1}^{N}p_{\modeloutput_i'}=\prod_{i=1}^{M}p_{\modeloutput_i}\prod_{j=M+1}^{N}p_{\modeloutput_j'}\\
    &\leq\prod_{i=1}^{M}p_{\modeloutput_i}%\\
    \leq\prod_{j\in\{j|\modeloutput_j\neq1,j \leq M\}}p_{\modeloutput_j}.
\end{aligned}
\end{equation}

Then let $K=|\{i|\modeloutput_i\neq1,i \leq M\}|$, we can obtain

\begin{equation}
\prod_{j\in\{j|\modeloutput_j\neq1,j \leq M\}}p_{\modeloutput_j}
\leq \prod_{j\in\{j|\modeloutput_j\neq1,j \leq M\}}p_2
= p_2^{K}
\leq \left(\frac{1}{2}\right)^{K}.
\end{equation}

Therefore, we get
\begin{equation}
\prob{\{\modeloutput_i'\}|\{\pair{\modelinput_i, \modelinput_i'}\}, \{\modeloutput_i\}, \studentmodel} \leq \left(\frac{1}{2}\right)^{K}.
\end{equation}

Finally, let $\left(\frac{1}{2}\right)^{K}\leq\alpha$, we can derive that
\begin{equation}
\begin{aligned}
K \geq \log_2 \frac{1}{\alpha}.
\end{aligned}
\label{equ:theroem1}
\end{equation}

That is, if $K \geq \ceil{\log_{2}\frac{1}{\alpha}}$ ($K$ is a positive integer), we can prove that $\prob{\{\modeloutput_i'\}|\{\pair{\modelinput_i, \modelinput_i'}\}, \{\modeloutput_i\}, \studentmodel} \leq \alpha $, i.e., reject the null hypothesis $H_0$ to conclude that $\studentmodel$ is not a random classifier, and hence continue further fingerprinting inference.

According to our definition of supporting set (\autoref{definition:supporting_set}), there exist two possible scenarios:
\begin{itemize}
\item Scenario 1: class 1 does not appear in the supporting set. By our result~\autoref{equ:theroem1}, if the size of the supporting set $|\supportset|$ is not smaller than $\ceil{\log_{2}\frac{1}{\alpha}}$, the attacker will reject $H_0$ and continue further inference.
\item Scenario 2: class 1 appears in the supporting set. Suppose class $k$ is out of the supporting set, then we have:
\begin{equation}
\begin{aligned}
K&=|\supportset|-|\{i|y_i==1,i \leq M\}|+|\{j|y_j==k, j \leq M\}|\\
 &\geq |\supportset| \geq \ceil{\log_{2}\frac{1}{\alpha}}.
\end{aligned}
\end{equation}
The first inequality is based on that fact that $|\{j|y_j==k, j \leq M\}|\geq|\{i|y_i==1,i \leq M\}|$.
\end{itemize}

Therefore, when the size of the supporting set satisfies $|\supportset| \geq \ceil{\log_{2}\frac{1}{\alpha}}$, we will reject the null hypothesis $H_0$ and continue further fingerprinting inference.
\end{proof}

\begin{remark}
\autoref{equ:theroem1} gives an easy to calculate but pretty strict condition.
In fact, we can simply induce a more precise one to reject $H_0$:
\begin{equation}
\begin{aligned}
    \prod_{j\in\{j|\modeloutput_j\neq1,j \leq M\}}p_{\modeloutput_j}
    &\leq \prod_{i=2}^{c}\left(\frac{1}{i}\right)^{|\{j|y_j==i,j \leq M\}|}
    \leq \alpha.
\end{aligned}
\label{equ:precise_condition}
\end{equation}
In the supporting set, we firstly reassign the class label $\{y_i\}$ from the most frequent class to the least frequent class with 2 to $c$, and then check the condition \autoref{equ:precise_condition}.  
\end{remark}

%==================================================
\section{Attack Performance with MNIST, CelebA and Random Noise as Probing Dataset}
%================================================== 

\begin{figure}[!htb]
\centering
\includegraphics[width=1\columnwidth]{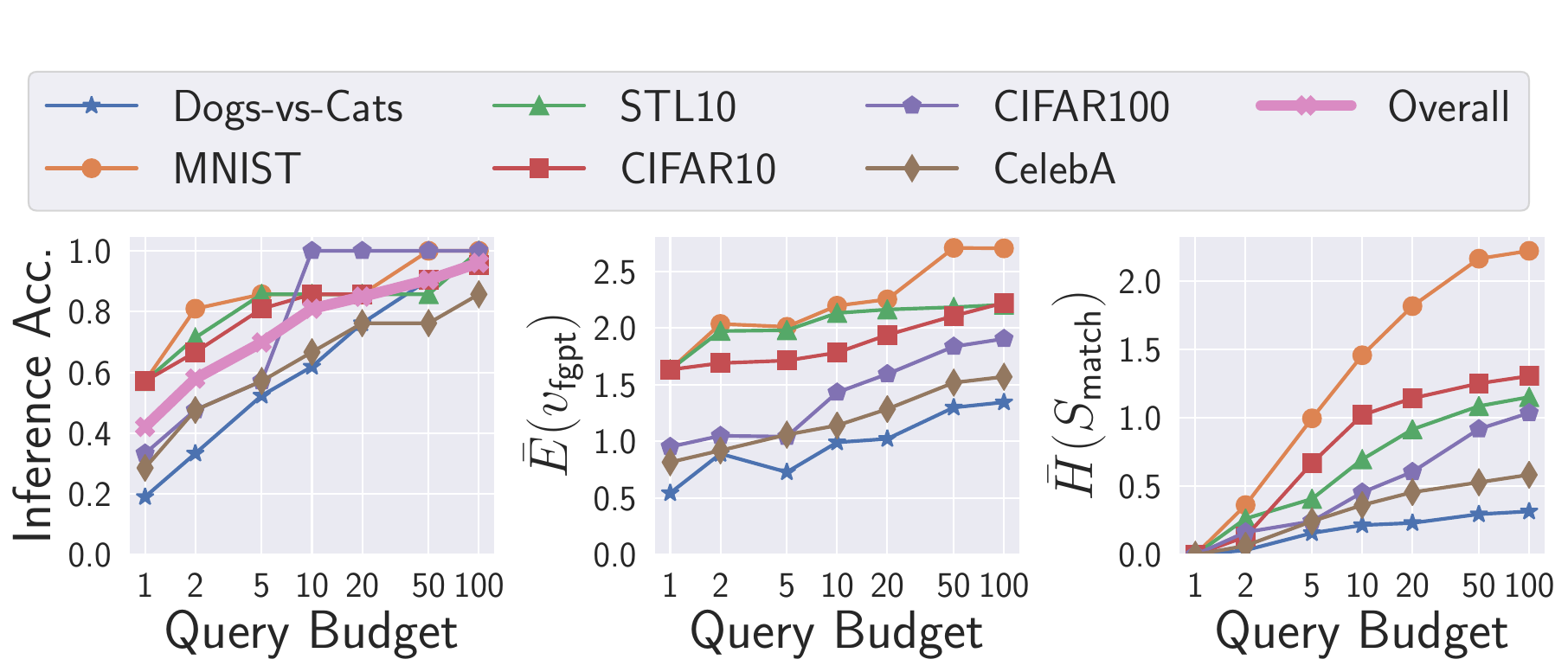}
\caption{
Performance with MNIST images as probing inputs. We have achieved a high average eccentricity of $\fingerprintingvector$ and a high average entropy of $\matchset$ on the MNIST classification task.
}
\label{fig:performance_with_MNIST}
\end{figure}

\begin{figure}[!htb]
\centering
\includegraphics[width=1\columnwidth]{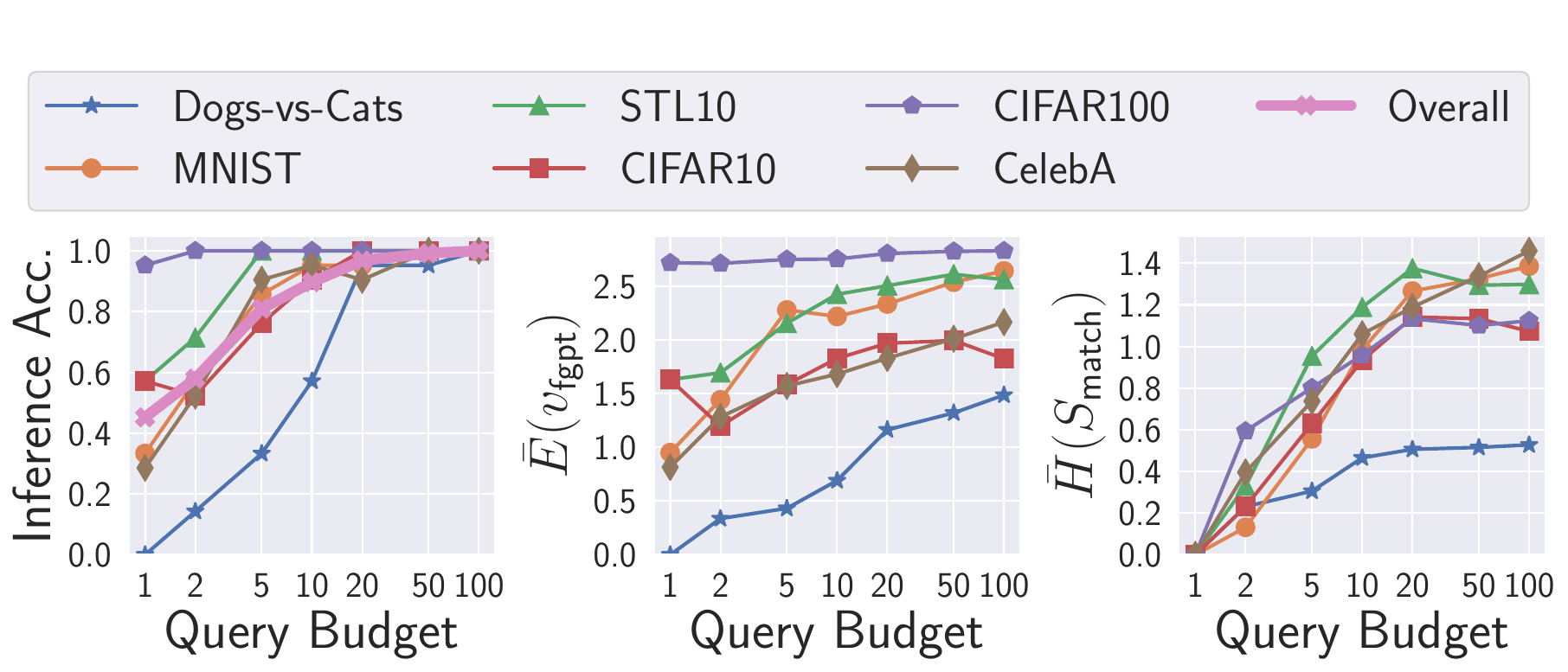}
\caption{
Performance with CelebA images as probing inputs. Compared to~\autoref{fig:performance_with_MNIST}, the three heuristics obviously increase on the CelebA classification task.
} 
\label{fig:performance_with_celeba}
\end{figure}

\begin{figure}[!htb]
\centering
\includegraphics[width=1\columnwidth]{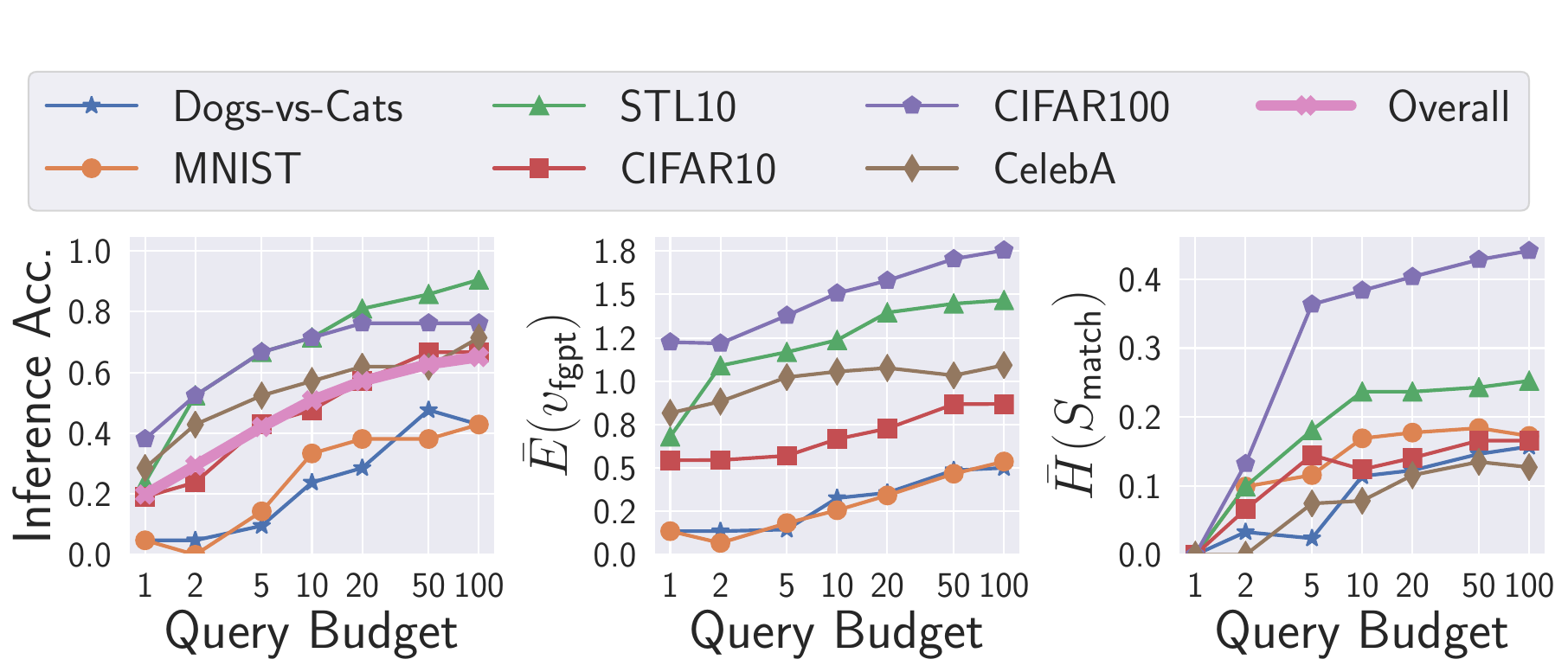}
\caption{
Performance with Random Noise as probing inputs. 
We have got the worst results on the three heuristics.
}
\label{fig:performance_with_noise}
\end{figure}

As a supplementary to~\autoref{section:inference_resource}, we plot the attack performance with MNIST, CelebA and Random Noise as the attack probing dataset in~\autoref{fig:performance_with_MNIST}, \autoref{fig:performance_with_celeba} and \autoref{fig:performance_with_noise}. 

%================================================== 
\section{Teacher Model with Latent Backdoor}
\label{appendix:latent_backdoor}
%==================================================

\begin{figure}[H]
\centering
\includegraphics[width=.30\textwidth]{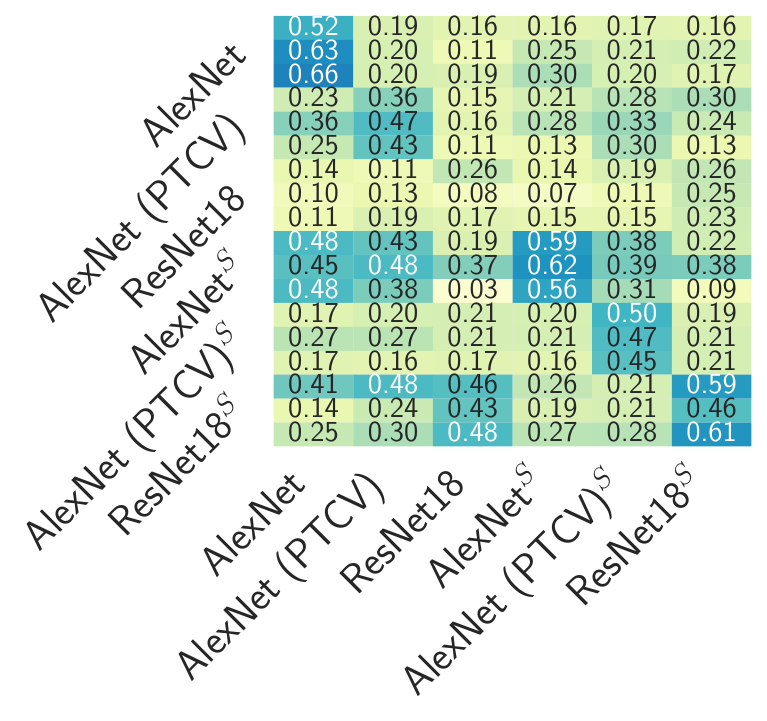}
\caption{
Teacher model fingerprinting vectors against clean and backdoored STL10 student models.
${ModelName}^{S}$ refers to the backdoored model for STL10 dataset.
}
\label{figure:stl10_backdoored}
\end{figure}

In our experiment, we separate CIFAR10, MNIST, and STL10 with the same ratio with the \texttt{Digit} attack in~\cite{YLZZ19}.
To achieve a more robust backdoor attack, we increase the number of target images to 200, and the trigger size is 40$\times$40 pixel.
We exhibit some attack results for the STL10 dataset in~\autoref{figure:stl10_backdoored}.

%==================================================
\section{Teacher Model Information}
%==================================================

In our experiments, we have downloaded pre-trained models from the PyTorch official repository, including AlexNet, DenseNet121, MobileNetV2, ResNet18, VGG16, VGG19, and GoogLeNet.
Besides, we have also downloaded another pre-trained AlexNet from the PyTorchCV repository to elaborate that our proposed attack can discriminate different teacher models with the same model architecture.
The sources of our obtained teacher models are listed in~\autoref{table:teacher_model}.

\begin{table*}[t]
\centering
\scriptsize
\setlength\tabcolsep{1.5pt}
\begin{tabularx}{\textwidth}{l | l | l | l | l | X | l}
\toprule
\textbf{Model} & \textbf{Provider} & $\teacherdataset$ & \textbf{\#BLKs} & \textbf{Candidate?} & \textbf{URL} & \textbf{MD5 Checksum}\\
\midrule
AlexNet & PyTorch & ImageNet & 5 & \cmark & \url{https://download.pytorch.org/models/alexnet-owt-4df8aa71.pth} & aed0662f397a0507305ac94ea5519309\\
\makecell[l]{AlexNet\\(PTCV)} & PyTorchCV & ImageNet & 5 & \cmark & \url{https://github.com/osmr/imgclsmob/releases/download/v0.0.384/alexnetb-1900-55176c6a.pth.zip} & 07e23324e570d9e1f10e280a53c509e3\\
DenseNet121 & PyTorch & ImageNet & 11 & \cmark & \url{https://download.pytorch.org/models/densenet121-a639ec97.pth} & a7047f0b44515469c3965e85cc31e512\\
MobileNetV2 & PyTorch & ImageNet & 18 & \cmark & \url{https://download.pytorch.org/models/mobilenet_v2-b0353104.pth} & f20b50b44fdef367a225d41f747a0963\\
ResNet18 & PyTorch & ImageNet & 5 & \cmark & \url{https://download.pytorch.org/models/resnet18-5c106cde.pth} & e5f5fcaec0feff7287f61b7bf461e8b2\\
VGG16 & PyTorch & ImageNet & 6 & \cmark & \url{https://download.pytorch.org/models/vgg16-397923af.pth} & 463aeb51ba5e122501bd03f4ad6d5374\\
VGG19 & PyTorch & ImageNet & 6 & \cmark & \url{https://download.pytorch.org/models/vgg19-dcbb9e9d.pth} & 92881fe292bd7d2408ecff58a101fd03\\
GoogLeNet & PyTorch & ImageNet & 10 & \xmark & \url{https://download.pytorch.org/models/GoogLeNet-1378be20.pth} & beba28483167f26c03ed26a6b569bbf9\\
\bottomrule
\end{tabularx}
\caption{
Sources of the teacher models used in our experiments.
}
\label{table:teacher_model}
\end{table*}

%================================================== 
\section{Transfer Learning Setup}
\label{sec:transfer_learning_setup}
%==================================================

The basic architecture of our black-box student model is one pre-trained feature extractor concatenated with fully connected layers.
We adopt the whole convolution part of each pre-trained model listed in~\autoref{table:teacher_model} as the feature extractor $\featureextractor$, and then concatenate it with two or three fully connected layers.
Particularly, for VGG16 and VGG19, we also adopt components except of the last layer from the pre-trained fully connected layers, which constitute the last teacher feature extractor block.
For each feature extractor, we build up three different student models individually to perform a more comprehensive evaluation. 
For single-label classification tasks, we use the Softmax activation function at the student model output.
While for the multilabel classification task on CelebA dataset, we use the Sigmoid activation function at the student model output.
\autoref{table:student_model} reports the student model architectures.

In our experiments, we use the Adam optimizer~\cite{KB15} to train the student models. 
When the number of fine-tuned blocks is no larger than three we set the learning rate to $10^{-3}$, and otherwise, we set the learning rate to $10^{-5}$.
\autoref{table:transfer_learning_acc} reports the average testing accuracy of the victim student models.

\begin{table}[!t]
\centering
\begin{threeparttable}
\scriptsize
\setlength\tabcolsep{1.5pt}
\begin{tabularx}{.46\textwidth}{ c | c }
\toprule
$\studentdataset$  & \makecell{\textbf{Model Architecture}*} \\
\midrule
Dogs-vs-Cats & 
\makecell[l]{
$\featureextractor$ $\rightarrow$ \fullyconnected{128} $\rightarrow$ \batchnorm $\rightarrow$ \softmax{2} \\
$\featureextractor$ $\rightarrow$ \fullyconnected{512} $\rightarrow$ \batchnorm $\rightarrow$ \fullyconnected{128} $\rightarrow$ \batchnorm $\rightarrow$ \softmax{2} \\
$\featureextractor$ $\rightarrow$ \fullyconnected{1024} $\rightarrow$ \batchnorm $\rightarrow$ \fullyconnected{256} $\rightarrow$ \batchnorm $\rightarrow$ \softmax{2}
} \\
\midrule
\makecell{MNIST,\\ STL10,\\ CIFAR10} & 
\makecell[l]{
$\featureextractor$ $\rightarrow$ \fullyconnected{128} $\rightarrow$ \batchnorm $\rightarrow$ \softmax{10} \\
$\featureextractor$ $\rightarrow$ \fullyconnected{512} $\rightarrow$ \batchnorm $\rightarrow$ \fullyconnected{128} $\rightarrow$ \batchnorm  $\rightarrow$ \softmax{10} \\
$\featureextractor$ $\rightarrow$ \fullyconnected{1024} $\rightarrow$ \batchnorm $\rightarrow$ \fullyconnected{256} $\rightarrow$ \batchnorm  $\rightarrow$ \softmax{10}
} \\
\midrule
CIFAR100 & 
\makecell[l]{
$\featureextractor$ $\rightarrow$ \fullyconnected{512} $\rightarrow$ \batchnorm $\rightarrow$ \softmax{100} \\
$\featureextractor$ $\rightarrow$ \fullyconnected{1024} $\rightarrow$ \batchnorm $\rightarrow$ \fullyconnected{256} $\rightarrow$ \batchnorm $\rightarrow$ \softmax{100} \\
$\featureextractor$ $\rightarrow$ \fullyconnected{4096} $\rightarrow$ \batchnorm $\rightarrow$ \fullyconnected{512} $\rightarrow$ \batchnorm $\rightarrow$ \softmax{100}
} \\
\midrule
CelebA & 
\makecell[l]{
$\featureextractor$ $\rightarrow$ \fullyconnected{256} $\rightarrow$ \batchnorm $\rightarrow$ \sigmoid{40} \\
$\featureextractor$ $\rightarrow$ \fullyconnected{1024} $\rightarrow$ \batchnorm $\rightarrow$ \fullyconnected{256} $\rightarrow$ \batchnorm  $\rightarrow$ \sigmoid{40} \\
$\featureextractor$ $\rightarrow$ \fullyconnected{2048} $\rightarrow$ \batchnorm $\rightarrow$ \fullyconnected{512} $\rightarrow$ \batchnorm  $\rightarrow$ \sigmoid{40}} \\
\bottomrule
\end{tabularx}
\begin{tablenotes}
   \item[*] For simplicity, \fullyconnected{$n$} refers to a fully connected layer with $n$ neuron.
   In our experiment, we choose the ReLU activation for fully connected layers, where are all followed by a dropout layer with rate 0.5. 
   \softmax{$n$} refers to a Softmax layer with $n$ outputs, \sigmoid{$n$} refers to a Sigmoid layer with $n$ outputs, and $\batchnorm$ refers to a batch normalization layer.
\end{tablenotes}
\end{threeparttable}
\caption{
Student model architectures used in our experiments.
}
\label{table:student_model}
\end{table}

\begin{table}[!t]
\centering
\scriptsize
\begin{threeparttable}
\setlength\tabcolsep{1.5pt}
\begin{tabular}{c|c|c|c|c|c|c|c|c}
\toprule
\multirow{2}{*}{\makecell{\textbf{Fine}\\\textbf{tuning}*}} & \multirow{2}{*}{\makecell{\\$\studentdataset$}} & \multicolumn{7}{c}{\textbf{Teacher model}}\\
\cline{3-9}
& & AlexNet & \makecell{AlexNet\\(PTCV)} & \makecell{Dense-\\Net121} & \makecell{Mobile-\\NetV2} & ResNet18 & VGG16 & VGG19 \\
\midrule
\multirow{6}{*}{\makecell{Fixed}} 
& Dogs-vs-Cats & 95.17 & 95.90 & 99.00 & 98.49 & 98.73 & 98.91 & 98.89 \\ 
& MNIST & 99.28 & 99.30 & 99.29 & 99.13 & 98.14 & 96.86 & 96.40 \\ 
& STL10 & 75.55 & 84.14 & 96.80 & 95.05 & 94.95 & 92.27 & 92.25 \\ 
& CIFAR10 & 62.35 & 71.39 & 91.80 & 89.87 & 88.02 & 71.57 & 67.50 \\ 
& CIFAR100 & 30.71 & 40.30 & 68.42 & 64.17 & 60.52 & 29.30 & 32.05 \\ 
& CelebA & 87.65 & 87.95 & 88.27 & 87.73 & 86.33 & 85.59 & 85.62 \\ 
\midrule
\multirow{6}{*}{\makecell{Last\\BLK}} 
& Dogs-vs-Cats & 96.66 & 97.09 & 98.65 & 98.21 & 98.33 & 98.67 & 98.73 \\ 
& MNIST & 99.43 & 99.47 & 99.21 & 98.98 & 99.38 & 99.09 & 99.10 \\ 
& STL10 & 88.11 & 88.37 & 96.03 & 94.20 & 92.04 & 93.46 & 94.08 \\ 
& CIFAR10 & 87.35 & 88.20 & 87.79 & 84.16 & 89.55 & 87.24 & 87.13 \\ 
& CIFAR100 & 57.86 & 59.45 & 62.64 & 54.89 & 59.95 & 57.28 & 57.27 \\ 
& CelebA & 88.36 & 88.49 & 88.57 & 87.90 & 88.02 & 86.92 & 86.84 \\ 
\midrule
\multirow{6}{*}{\makecell{Last\\ two \\BLKs}} 
& Dogs-vs-Cats & 96.48 & 96.46 & 98.95 & 98.38 & 97.65 & 98.38 & 98.53 \\ 
& MNIST & 99.48 & 99.53 & 99.17 & 99.13 & 99.26 & 99.37 & 99.37 \\ 
& STL10 & 86.10 & 85.40 & 95.34 & 94.03 & 86.37 & 91.08 & 91.28 \\ 
& CIFAR10 & 87.69 & 88.28 & 90.22 & 85.84 & 89.30 & 89.17 & 88.10 \\ 
& CIFAR100 & 58.32 & 56.50 & 65.70 & 56.55 & 54.52 & 48.02 & 41.71 \\ 
& CelebA & 88.81 & 88.59 & 88.41 & 88.13 & 88.33 & 88.01 & 88.00 \\  
\midrule
\multirow{6}{*}{\makecell{Last\\ three \\BLKs}} 
& Dogs-vs-Cats & 95.07 & 95.53 & 98.83 & 98.33 & 96.88 & 98.13 & 96.27 \\ 
& MNIST & 99.52 & 69.57 & 99.20 & 99.24 & 99.42 & 99.10 & 99.41 \\ 
& STL10 & 80.59 & 79.69 & 95.27 & 93.01 & 84.23 & 90.12 & 87.15 \\ 
& CIFAR10 & 86.32 & 85.52 & 91.22 & 87.30 & 89.02 & 84.53 & 80.84 \\ 
& CIFAR100 & 50.42 & 48.51 & 66.73 & 58.06 & 49.75 & 44.35 & 40.92 \\ 
& CelebA & 88.51 & 88.34 & 88.45 & 88.28 & 88.27 & 88.03 & 77.47 \\  
\midrule
\multirow{6}{*}{\makecell{Last\\ four \\BLKs}} 
& Dogs-vs-Cats & 94.14 & 94.69 & 98.39 & 98.31 & 96.53 & 96.31 & 87.83 \\ 
& MNIST & 99.44 & 99.40 & 99.37 & 99.34 & 99.41 & 99.37 & 99.10 \\ 
& STL10 & 75.69 & 71.31 & 91.56 & 92.86 & 77.40 & 87.22 & 85.94 \\ 
& CIFAR10 & 86.29 & 86.37 & 91.55 & 87.64 & 89.07 & 76.10 & 66.99 \\ 
& CIFAR100 & 48.88 & 43.89 & 62.26 & 57.79 & 43.77 & 44.65 & 45.47 \\ 
& CelebA & 88.67 & 88.47 & 88.60 & 88.14 & 88.17 & 87.97 & 87.57 \\ 
\midrule
\multirow{6}{*}{\makecell{Last\\ five \\BLKs}} 
& Dogs-vs-Cats & 97.31 & 97.61 & 99.19 & 98.73 & 98.87 & 99.04 & 99.09 \\ 
& MNIST & 99.56 & 99.56 & 99.50 & 99.53 & 99.60 & 99.61 & 99.59 \\ 
& STL10 & 85.81 & 84.58 & 97.19 & 95.67 & 93.81 & 95.78 & 95.83 \\ 
& CIFAR10 & 89.77 & 90.51 & 94.70 & 89.99 & 94.67 & 93.51 & 93.48 \\ 
& CIFAR100 & 34.30 & 35.33 & 51.31 & 40.92 & 39.86 & 44.35 & 48.55 \\ 
& CelebA & 88.44 & 88.52 & 89.07 & 88.59 & 88.49 & 88.91 & 88.85 \\  
\bottomrule
\end{tabular}
\begin{tablenotes}
   \item[*]For simplicity, ``Fixed'' refers to that pre-trained parameters are fixed, ``Last BLK'' refers to that the last block of the teacher feature extractor gets fine-tuned, ``Last two BLKs'' refers to that the two last blocks of the teacher feature extractor get fine-tuned, and so on.
\end{tablenotes}
\end{threeparttable}
\caption{
Average testing accuracy (\%) of the victim student models in our experiments.
}
\label{table:transfer_learning_acc}
\end{table}

%================================================== 
\section{Fingerprinting Vectors for Fine-tuned Student Models}
%==================================================

\begin{figure*}[!th]
\centering
\includegraphics[width=1.7\columnwidth]{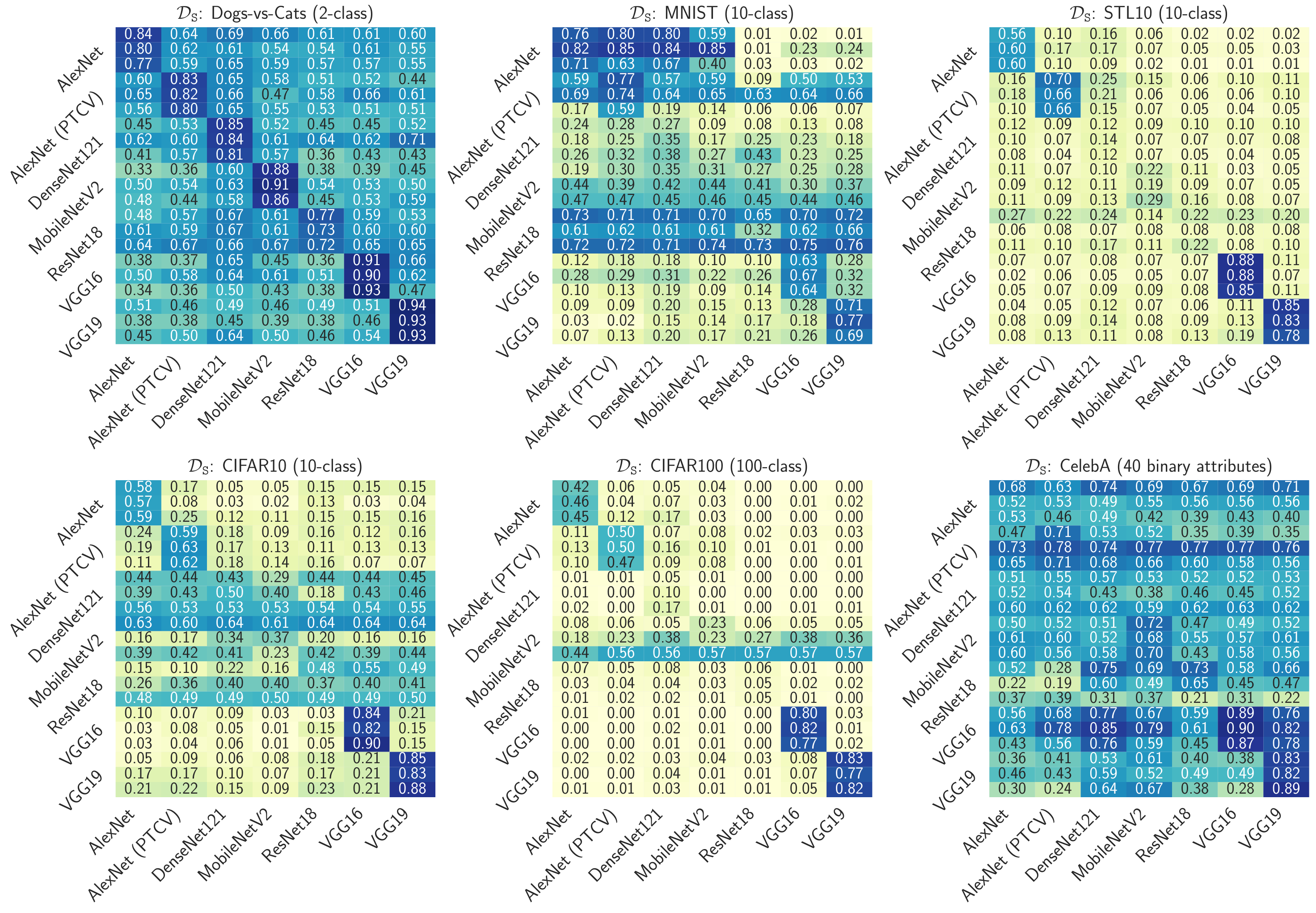}
\caption{
Teacher model fingerprinting vectors w.r.t.\ different classification tasks (100 fingerprinting pairs for each teacher model candidate, when the last block gets fine-tuned).
}
\label{fig:last_one_fine_tuned}
\end{figure*}

\begin{figure*}[!th]
\centering
\includegraphics[width=1.7\columnwidth]{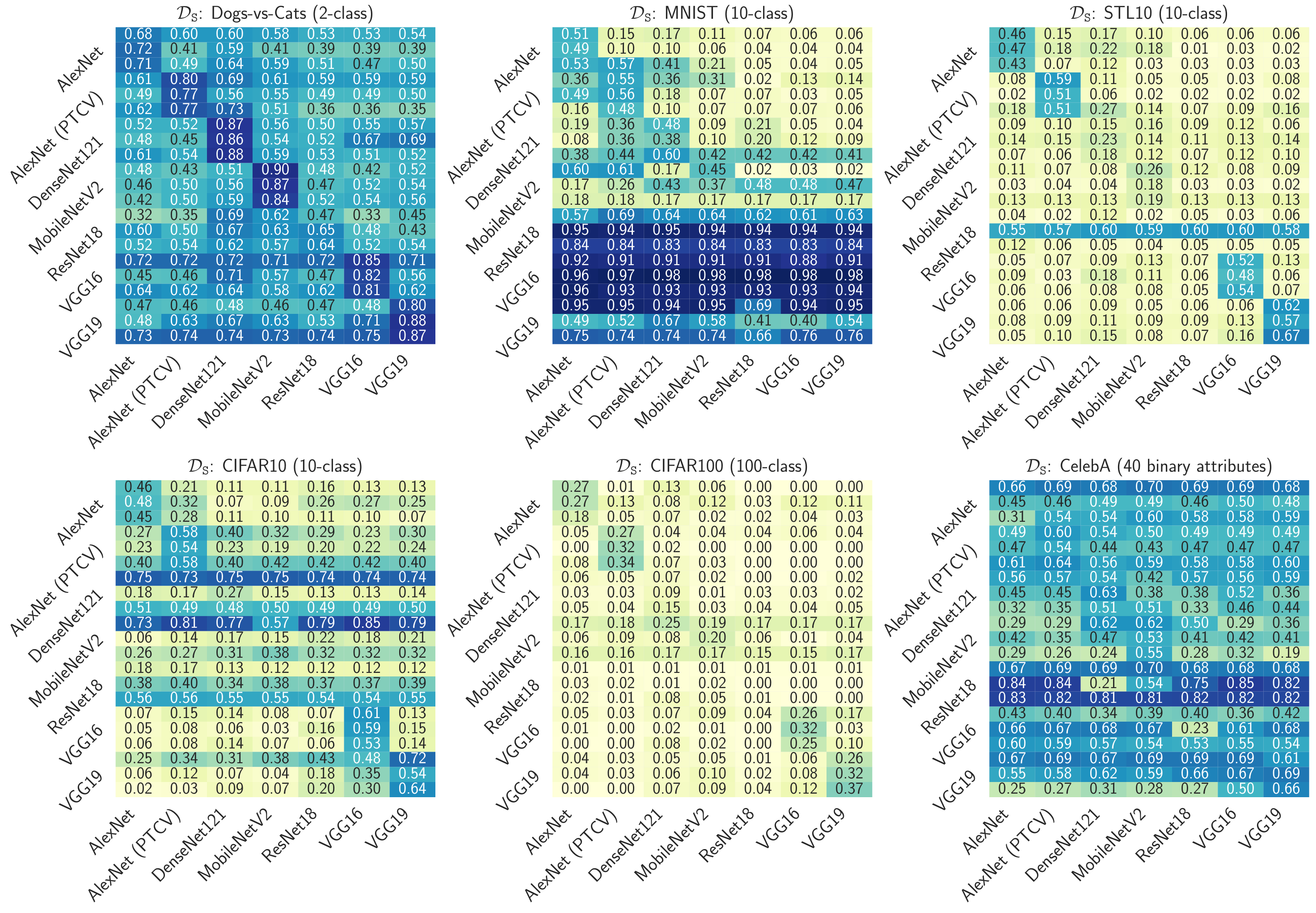}
\caption{
Teacher model fingerprinting vectors w.r.t.\ different classification tasks (100 fingerprinting pairs for each teacher model candidate, when the last two blocks get fine-tuned).
}
\label{fig:last_two_fine_tuned}
\end{figure*}

In this part, we exhibit fingerprinting vectors when the last block gets fine-tuned and the last two blocks get fine-tuned in~\autoref{fig:last_one_fine_tuned} and~\autoref{fig:last_two_fine_tuned}, respectively.

%==================================================
\end{document}